\documentclass[a4paper,11pt]{article}

\usepackage[a4paper,hmargin=2.8cm,vmargin=3cm]{geometry}
\usepackage[english]{babel}
\usepackage[intlimits]{amsmath}
\usepackage{amsthm,amsfonts,amssymb}
\usepackage{mathtools}
\usepackage{mathrsfs}
\usepackage{bbm}
\usepackage{authblk}
\usepackage{upgreek}
\usepackage{slashed}
\usepackage{graphicx}
\usepackage[dvipsnames]{xcolor}
\usepackage{tikz}
\usetikzlibrary{
    shapes.geometric,
    decorations.pathmorphing,
    decorations.pathreplacing,
    decorations.shapes,
    positioning
}

\usepackage{subcaption}
\usepackage{enumitem}
\usepackage{dsfont}
\definecolor{BeauBlue}{rgb}{0,0.2,0.85}
\definecolor{BeauOrange}{rgb}{0.8,0.1,0}

\usepackage{hyperref}

\hypersetup{
    colorlinks = true,
    linkcolor = BeauBlue,
    urlcolor  = BeauBlue,
    citecolor = BeauOrange,
}

\numberwithin{equation}{section}

\renewenvironment{thebibliography}[1]
{
    \begin{oldthebibliography}{#1}
    \setlength{\itemsep}{0em}
    \setlength{\parskip}{0.5em}
}
{
    \end{oldthebibliography}
}

\newcommand{\ee}{\mathrm{e}}
\newcommand{\ii}{\mathrm{i}}
\newcommand{\dd}{\mathrm{d}}
\newcommand{\veps}{\varepsilon}
\newcommand{\Do}{\mathcal{D}}
\renewcommand{\a}{\alpha}
\newcommand{\x}{\hat{x}}
\newcommand{\y}{\hat{y}}

\newcommand\restr[2]{{
    \left.\kern-\nulldelimiterspace
    #1
    \vphantom{|}
    \right|_{#2}
}}

\renewcommand{\phi}{\varphi}

\newcommand{\GA}{\mathcal{A}}
\renewcommand{\d}[1]{\operatorname{d}\!#1}
\renewcommand{\Re}{\operatorname{Re}}
\renewcommand{\Im}{\operatorname{Im}}

\newcommand{\La}{\Lambda}
\newcommand{\LC}{|\Lambda|}

\newcommand{\tr}[1]{\operatorname{Tr}\,#1}

\newcommand{\Z}{\mathbb{Z}}

\newcommand{\E}{\mathbb{E}}
\newcommand{\C}{\mathbb{C}}
\newcommand{\R}{\mathbb{R}}
\newcommand{\N}{\mathbb{N}}

\newcommand{\ov}[1]{\overline{#1}}

\newcommand{\dPhi}[1]{\d{\Phi}_{#1}\,}
\newcommand{\dphi}[1]{\d{\phi}_{#1}\,}

\theoremstyle{definition}
\newtheorem{definition}{Definition}[section]
\newtheorem{remark}[definition]{Remark}

\theoremstyle{plain}
\newtheorem{theorem}[definition]{Theorem}
\newtheorem{proposition}[definition]{Proposition}
\newtheorem{lemma}[definition]{Lemma}

\title{Density of states for random Schr\"odinger operators with linearly correlated heavy-tailed disorder}
\author[1]{Margherita Disertori}
\author[2]{Luca Fresta}
\author[1]{Liza Schonlau}
\affil[1]{Hausdorff Center for Mathematics,
University of Bonn, Endenicher Allee 60
53115 Bonn, Germany\\Email: \href{mailto: mdiserto@uni-bonn.de}
{\texttt{mdiserto@uni-bonn.de}},
\href{mailto: aschonla@uni-bonn.de}
{\texttt{aschonla@uni-bonn.de}
}}
\affil[2]{Mathematics Department, University Roma Tre,
Largo San Leonardo Murialdo 1, 00146 Roma, Italy\\Email: \href{mailto:luca.fresta@uniroma3.it}
{\texttt{luca.fresta@uniroma3.it}} }

\date{\today}

\begin{document}

\maketitle
\begin{abstract}
We study the density of states of random Schr\"odinger operators with linearly correlated heavy-tailed disorder. Using an integration-by-parts supersymmetric representation, we prove smoothness of the density of states and Wegner estimates in the strong-disorder regime. We are particularly interested in mixed-sign correlation kernels, which serve as a toy model for the hyperbolic structures arising in the analysis of averaged squared Green's functions. Our approach applies to a broad class of correlation kernels and disorder distributions, including $\alpha$-stable laws.
\end{abstract}

\tableofcontents

\section{Introduction}

Random Schr\"odinger operators were introduced by Anderson \cite{Anderson1958} to describe the metal-insulator transition in crystalline structures, with a random potential modelling the effect of impurities in the crystal. The conducting properties of the system are determined by the spectral properties of the associated Hamiltonian. The Anderson Hamiltonian on $\ell^{2}(\mathbb{Z}^{d})$ takes the form
\begin{equation}\label{eq:anderson}
H = -\Delta +  \lambda V \;,
\end{equation}
where $-\Delta $ is the lattice Laplacian $ (-\Delta u)(x) = \sum_{y\in\Z^d:\, \lvert y-x\rvert = 1} \left(u(x)-u(y)\right)$, $
\lambda \in \mathbb{R}$ is the disorder strength, and $V$ is a random multiplicative potential $(Vu)(x) = V_x u(x)$, with $(V_{x})_{x \in \Z^{d}}$ i.i.d.~centred random variables.
In the insulating phase, the spectrum is pure point and eigenfunctions are spatially localised. This regime has been rigorously understood starting with the pioneering works based on the multiscale analysis \cite{Fro83,Fro85} and the fractional-moment method \cite{Aiz93}. These techniques have been extensively developed in subsequent works, see, e.g., \cite{Carmona1987,Bourg,Elgart,Ding20}. Alternative approaches based on supersymmetry have also been explored, see \cite{klein86,Wang,Fresta2021}.

In contrast, the metallic phase is much less understood. It is expected that in three or higher dimensions and for sufficiently weak disorder, random Schr\"odinger operators exhibit absolutely continuous spectral components associated with extended states and nontrivial transport \cite{Anderson79}. However, rigorous results in this direction are limited to effective models \cite{Klein,Froe07}. We refer to \cite{KirschInvitation2007} for an introduction and to \cite{aizenman2015random} for a detailed survey. 
\smallskip

In addition to the metal-insulator transition, a central question concerns the density of states measure (DOS), which describes the distribution of the energy levels in the infinite-volume limit. Not only does the density of states enter many equilibrium quantities, but it also plays an important role for the transport properties \cite{Fro85,Klopp} and for spectral statistics, where it determines the natural local scale of eigenvalues in the thermodynamic limit.
Understanding the DOS for random Schr\"odinger operators in full generality remains a challenging problem. 
Most rigorous results concern the localisation regime, where regularity properties of the DOS have been extensively studied. In contrast, in the delocalisation regime, only partial results have been established through multi-scale analysis of hierarchical models \cite{Bovier1990,AFP20}, or in effective models like random band matrices \cite{Yau25,Erdos25} and supersymmetric nonlinear sigma models \cite{Disertori2010}.

For i.i.d.~strong disorder, a basic regularity result shows that the DOS is absolutely continuous \cite{Delyon1984}, and in general one typically only has log-H\"older continuity of the integrated density of states \cite{Craig1983}, so that the corresponding density may be unbounded. A classical Wegner estimate \cite{Wegner1979} implies a uniform bound on the density under the additional assumption of a bounded single-site density. Higher regularity requires stronger assumptions on the single-site distribution. If the single-site law is analytic in a complex strip, then the DOS is analytic at sufficiently large disorder or energy \cite{Constantinescu1984}. This analyticity assumption can be relaxed to exponential decay of the characteristic function \cite{Carmona1990}. In a different direction, Bovier--Campanino--Klein--Perez \cite{Bovier1988} used a supersymmetric formulation combined with a cluster expansion to prove smoothness of the DOS in the strong-disorder regime.
\bigskip

While most results concern i.i.d.~random potentials, physically relevant models often exhibit spatial correlations. However, substantially less is known for correlated disorder: correlations may introduce severe technical difficulties, and even basic ergodic properties can become non-trivial in the general case \cite{Combes98}.
In this work, we consider a class of random Schr\"odinger operators on $\ell^2(\Z^d)$ with linearly correlated potentials of the form
\begin{equation}\label{eq:definition_of_correlated_pot_in_intro}
V_{x}:=   \sum_{ y \in \mathbb{Z}^{d}} T_{x y} \; W_{y} \;,
\end{equation}
where $(W_y)_{y\in\Z^d}$ are centred i.i.d.~random variables and $T = (T_{x y})_{x,y \in \mathbb{Z}^{d}}$ is a symmetric kernel encoding the correlations. Linear correlations provide a natural intermediate class between independent and fully correlated disorder while retaining enough algebraic structure to remain analytically tractable.

We are particularly interested in mixed-sign kernels $T$, that is, kernels whose entries have indefinite sign. Our motivation comes from the supersymmetric analysis of correlation functions, where internal hyperbolic symmetry naturally emerge in the study of the averaged squared Green's function, leading to a behaviour analogous to mixed-sign kernels.
Controlling correlation functions with hyperbolic structure remains an open problem beyond one-dimensional systems and results are available only for effective models, see \cite{klein86,Disertori2010,Disertori2010+,Bao18,Shcherbina21}.
Although the present work does not reach the level of indefiniteness expected in those problems, we believe that studying such extensions is already of independent technical interest.

A second contribution of this work is an integration-by-parts supersymmetric representation obtained by performing the integration by parts at an early stage of the supersymmetric formalism. 
This avoids assuming boundedness of the moments of the disorder distribution and yields a representation that applies to a broad class of heavy-tailed laws, including $\alpha$-stable distributions. While related integration-by-parts arguments already appear implicitly in \cite{Bovier1988} at a later stage of the analysis, we formulate the method directly at the level of the supersymmetric representation and work throughout the paper in this general framework.

\subsection{Main result}
\label{sec:main-result}

Let $\Lambda\subset\mathbb Z^d$ be a finite cube and let $H_\Lambda$ denote the restriction of $H$ defined in \eqref{eq:anderson}. We take the random potential $V$ as in \eqref{eq:definition_of_correlated_pot_in_intro}, with $(W_y)_{y\in\Z^d}$ i.i.d., with characteristic function
\begin{equation}\label{eq:nu-0-def}
\nu_{0}(t) := \mathbb{E} \big[ \ee^{-\ii t W_{y}} \big] \;,\qquad \forall y \in \Z^{d} \;.
\end{equation}
Since the correlation kernel $T$ is not assumed to have finite range, the random variables $(V_x)_{x\in\Lambda}$ generally depend on the entire family $(W_y)_{y\in\Z^d}$. Accordingly, the series defining $V_x$ in \eqref{eq:definition_of_correlated_pot_in_intro} is not a priori well-defined; the precise assumptions ensuring its existence are stated in Theorem~\ref{thm:main}.

The finite-volume DOS of $H$ is the empirical measure associated with $H_{\Lambda}$
\begin{equation}\label{eq: empirical-density}
\mu_{\Lambda}:=\frac{1}{|\Lambda|} \sum_{e \in \mathrm{spec}(H_{\Lambda})} \delta_{e}\;,
\end{equation}
where $\delta_{e}$ denotes the Dirac point mass at $e \in \mathbb{R}$, and the sum runs over all the eigenvalues of $H_{\Lambda}$, counted with multiplicity. 
Given ergodicity, $\mu_\Lambda$ converges weakly (in probability) as $\Lambda \nearrow \mathbb{Z}^{d}$ to a deterministic measure $\mu$, which is the DOS of $H$~\cite{Figotin1992}.
In our setting, ergodicity is guaranteed whenever $T$ is periodic, namely $Q \Z^d$-translation invariant for some $Q \in \N$, meaning that
\begin{equation*}\label{eq:T-translation}
T_{x(y+Qz)} = T_{(x-Qz)y} \;,\qquad \forall x,y,z\in\Z^d \;.
\end{equation*}
We use the notation $a \vee b := \max(a,b)$ and $a \wedge b := \min(a,b)$. It will be convenient to decompose $T$ as follows
\begin{equation}\label{eq:T+T-}
T = T^+ - T^-,
\qquad
T^\pm_{xy} := (\pm T_{xy}) \vee 0 \;,
\end{equation}
and to consider the restriction of $T$ and $T^{\pm}$ on any subset $Y \subset \Z^{d}$. For $\alpha >0$, we equip the row-vectors $(T_{x \cdot})_{x \in Y}$ with the quasi-norm\footnote{This is a norm when $\a \geq 1$.}
\begin{equation}\label{eq:a-norm}
\| T_{x \cdot} \|_{Y,\a} := \Big(\sum_{y \in Y} |T_{xy}|^{\a} \Big)^{1/\a} \;,
\end{equation}
and similarly for $(T^{\pm}_{x \cdot})_{x \in Y}$.
Our main result establishes absolute continuity and smoothness of the DOS for a broad class of linearly correlated random potentials.
\begin{theorem}\label{thm:main}
Let $T = (T_{x y})_{x,y \in \mathbb{Z}^{d}}$ be a symmetric and periodic kernel.
Let $(W_y)_{y\in\Z^d}$ be centred i.i.d.~random variables with  characteristic function $\nu_{0}$. Fix $\a \in (0,2]$ and assume that
\begin{equation}\label{eq:beta-stable-tail-1}
|1-\nu_{0}(t)| \lesssim |t|^{\a}\;, \qquad \qquad \sup_{x \in \Z^{d}} \| T_{x \cdot}\|_{\Z^{d},\alpha} < \infty \;.
\end{equation}
Then, the random field
\begin{equation}\label{eq:random-V-theorem}
V_{x}:= \sum_{ y \in \mathbb{Z}^{d}} T_{x y} \; W_{y}
\end{equation}
is well-defined. 
Furthermore, assume that 
\begin{equation}\label{eq:beta-stable-tail-}
|\nu_{0}(t)| \leq \ee^{-|t|^{\a}} \;,
\end{equation}
and that either
\begin{equation}\label{eq:structural-T}
\begin{split}
& \inf_{Y \subset \Z^{d}} \Big(
\kappa_{\a} \inf_{x \in Y}\| T^{+}_{x \cdot} \|_{Y,1 \vee \a} - \sup_{x \in Y}\| T^{-}_{x \cdot}\|_{Y,1 \wedge \a}\Big) >0 \;, 
\\
 \mathrm{or} \qquad
& \inf_{Y \subset \Z^{d}} \Big(\kappa_{\a} \inf_{x \in Y}\| T^{-}_{x \cdot}\|_{Y,1 \vee \a} - \sup_{x \in Y}\| T^{+}_{x \cdot}\|_{Y,1 \wedge \a}\Big) >0  \;,
\end{split}
\end{equation}
where $\kappa_{\a}:= \mathbf{1}_{0 \leq \a \leq 1} + \mathbf{1}_{\a \geq 1} 2^{\frac{1}{\a}-1}$.
Then, there exists $\lambda_0 = \lambda_{0}(T,\alpha) <\infty$ such that for every
$\lambda \geq \lambda_0$ the density of states measure of
$H=-\Delta+\lambda V$ is Lebesgue absolutely continuous with $C^\infty(\mathbb R)$ density $\rho$ such that
\begin{equation*}
	\sup_{E\in\R}\left\lvert \partial^{k}_{E} \rho(E)\right\rvert \leq \frac{c_{k}}{\lambda^{k+1}} , \qquad\forall k\in \N_0,
\end{equation*}
for suitable constants $c_k$.
\end{theorem}
\begin{remark}\label{rmk:assumption-main}
Let us briefly comment on the assumptions.
\begin{itemize}
\item[i)] The assumptions in \eqref{eq:beta-stable-tail-1} ensure that the finite-dimensional truncations of $V$ converge.
The existence of the random field $V$ then follows from
L\'evy's continuity theorem, see Lemma \ref{lemma:exist-charact}.

\item[ii)] The distributions that we consider are characterised by the fact that their characteristic functions have at least sub-exponential decay. These include the $\a$-stable distributions $\nu_{0}(t) = \ee^{-|t|^{\a}} $. Note that if we required instead $|\nu_{0}(t)| = \ee^{-c|t|^{\a}}$, for some $c>0$, the latter constant could be absorbed into the disorder strength $\lambda$ by simple rescaling. Similarly, we considered $\lambda >0$ since the sign can be absorbed into the kernels $T$.

\item[iii)] The proof relies on moment bounds on the
finite-volume characteristic functions of $V$.
The structural condition \eqref{eq:structural-T} on $T$ guarantees the coercive estimate
\[
|\nu_{\Lambda}\big( r \big) |
\leq \exp\!\Big(-C\sum_{x\in \Lambda} r_x^\alpha\Big), \qquad r=(r_{x})_{x \in \Lambda} \in \R_{+}^{\Lambda} \;,
\]
for all $\Lambda \subset \mathbb{Z}^{d}$, with $\nu_{\Lambda}$ the finite-volume characteristic function of $V$ defined in \eqref{eq:nu-lambda}. This is crucial to control the supersymmetric integrals; see Theorem \ref{thm:SUSY_formula_averaged_Greens_function} and  Section~\ref{section:application}.

\item[iv)] One can easily exhibit examples of $T$ satisfying the structural condition \eqref{eq:structural-T}. For example, $T = \mathbbm{1} - \delta P$, where $P$ is the adjacency matrix of $\Z^{d}$. Then, one can check that
\begin{equation}
\inf_{x \in Y} \| T^{+}_{x \cdot}\|_{Y,1 \lor \a} = 1 \;,
\qquad
\sup_{x \in Y} \| T^{-}_{x \cdot}\|_{Y,1 \lor \a} \leq (2d)^{1/(1 \lor \a)} \delta \;.
\end{equation}
so that the first condition in \eqref{eq:structural-T} is satisfied if $\delta $ is small enough. Note that this example generalises the case studied in \cite{Disertori2020} to the case of negative correlations between all nearest-neighbour sites.

\end{itemize}
\end{remark}
\begin{remark}\label{rmk:comparison}
Theorem~\ref{thm:main} connects and extends two independent lines of results on the regularity of the DOS.
\begin{enumerate}
\item[i)] In the i.i.d.~case, Bovier--Campanino--Klein--Perez~\cite[Theorem~1.1]{Bovier1988} proved $C^n$-regularity of the DOS at strong disorder under the integrability condition
\begin{equation*}\label{eq:integrability_cond}
t \mapsto (1+|t|^{d+n})\,\nu_0(t)\in L^1(\R).
\end{equation*}
Our method extends this result to the special case of block-diagonal correlation kernels $T$ (see Remark \ref{eq: block-case}). However, considering correlation kernels with infinite range requires stronger decay assumptions on the characteristic function, as stated in \eqref{eq:beta-stable-tail-}.

\item[ii)] Disertori--Lager~\cite{Disertori2020} derived an exact supersymmetric representation of the DOS for linearly correlated Cauchy disorder ($\alpha =1$), under a structural restriction on $T$ that allows for only one single pair of negative correlations. We extend this framework to a broader class of kernels $T$ and to general disorder distributions with sub-exponentially decaying characteristic functions.
\end{enumerate}
\end{remark}
We approach the DOS through the finite-volume Borel--Stieltjes transforms, which can be expressed in terms of the Green's functions as follows
\begin{equation}\label{eq:borel-stieltjes-transform}
m_{\Lambda}(z):=\int_{\mathbb R}\frac{\mu_\Lambda(\d E')}{z-E'}
= -\frac{1}{|\Lambda|} \tr G_{\Lambda}(z) \;,
\qquad \quad
G_{\Lambda}(z):=(H_{\Lambda}-z)^{-1} \;,\qquad \quad  \Im z >0 \;,
\end{equation}
where $\mu_{\Lambda}$ is the finite-volume DOS introduced in \eqref{eq: empirical-density}.
Weak convergence of the measures $\mu_{\Lambda}$ to $\mu$ implies pointwise convergence of the Borel--Stieltjes transforms $m_{\Lambda}(z) \to m(z) = \int_{\mathbb R}\frac{\mu(\d E')}{z-E'}$ in the upper half-plane. By ergodicity, $\mu$ is deterministic, so that we can write
\begin{equation}\label{eq:m-overline}
m(z)
= -\lim _{\Lambda \nearrow \Z^{d}} \frac{1}{|\Lambda|} \tr \mathbb E \big[ G_{\Lambda}(z) \big] =\lim _{\Lambda \nearrow \Z^{d}}  \overline{m}_{\Lambda}(z) \;,
\qquad \quad\Im z >0 \;,
\end{equation}
where $\overline{m}$ is the Borel--Stieltjes transform of the averaged finite-volume measure $\mathbb{E} \, \big[ \mu_{\Lambda} \big]$.
Standard results in Herglotz theory imply that $m$ determines the absolutely continuous part of $\mu$ through the boundary value formula
\begin{equation}\label{eq:link_between_finite_vol_DOS_and_av_tr_resolvent}
\frac{\dd \mu^{\mathrm{ac}}}{\d E}(E)
= \frac{1}{\pi}\Im m(E+\ii 0^{+}) \;,
\end{equation}
whereas the singular part of $\mu$ is supported on the set where the limit is infinite. Consequently, uniform estimates on the averaged finite-volume Green's function imply estimates on the DOS. The same strategy applies to the derivatives in $E$, with the corresponding uniform bounds on the derivatives of the Green's functions.

To study the averaged Green's function, we use the supersymmetric (SUSY) approach introduced by Efetov et al.~in \cite{Efetov1980}, building on an earlier work by Wegner \cite{Wegner1979}.
This method represents averages of Green's functions $G_{\Lambda}(z)$, $z \in \mathbb{C} \setminus \mathbb{R}$, as oscillatory integrals involving complex and Grassmann variables. 
Let $(\ov{\xi}_{x},\xi_x)_{x \in \Lambda}$ be independent Grassmann variables, and $(\Phi_{x}^{*},\Phi_{x})_{x \in \Lambda}$ be supervectors of the form $\Phi_{x} = (\varphi_{x},\xi_{x})^{t}$, $\Phi^{*}= (\ov{\varphi}_{x},\ov{\xi}_{x})$, where $\varphi_{x}$ and $\ov{\varphi}_{x}$ are complex variables. If $M \in \C^{\Lambda \times \Lambda}$, we use the shorthand
\begin{equation*}
\Phi^{*} M \Phi := \sum_{x ,y} M_{xy} \Phi^{*}_{x} \Phi_{y}= 
\sum_{x ,y} M_{xy} \big(\ov{\varphi}_{x} \varphi_{y} + \ov{\xi}_{x} \xi_{y}  \big) \;,
\end{equation*}
see Appendix~\ref{appendix:GA} for details.
The supersymmetric method is rooted in the following identity 
\begin{equation}\label{eq:trace_as_susy_integral}
\big(G_\Lambda(E+\ii \varepsilon)\big)_{xy}
		= \ii\int \dPhi{\Lambda} \ee^{\ii\Phi^*(E+i\varepsilon+\Delta- \lambda V)\Phi} \,\xi_{x}\ov{\xi}_{y} \;,
\end{equation}
where $\dd \Phi_{\Lambda}$ denotes the flat Berezin superintegration form
\begin{equation*}
\dd \Phi_{\Lambda} = \prod_{x \in \Lambda} \frac{\dd \Re \phi_{x} \dd \Im \phi_{x}}{\pi} \partial_{\bar \xi_x} \partial_{ \xi_x} \;.
\end{equation*}
The only random term in (\ref{eq:trace_as_susy_integral}) is $\ee^{-\ii \lambda \Phi^* V\Phi}$, so that, assuming we can swap the disorder average with superintegration, we obtain the two-point function of an effective oscillatory supersymmetric integral with interaction $\mathbb{E} \big[ \ee^{-\ii \lambda \Phi^* V\Phi}\big]$. However, 
\begin{equation*}
\ee^{-\ii \lambda \Phi^* V\Phi} = \prod_{x \in \Lambda} \Big[ \ee^{-\ii \lambda V_x |\phi_x|^{2}}(1-\ii \lambda V_x \ov{\xi}_x\xi_x)\Big] \;,
\end{equation*}
so that making sense of $\mathbb{E}\big[ \ee^{-\ii \lambda \Phi^* V\Phi}\big]$ would at least require that $\mathbb{E}[|V_{x}|]$ is finite\footnote{For correlated potentials boundedness of higher moments is actually needed.}.

To circumvent this problem, in \cite{Disertori2020} they introduced a change of variables to supersymmetric polar coordinates $\Phi^* \mapsto \Phi^*_{\mathcal{S}}$ and $\Phi\mapsto\Phi_{\mathcal{S}}$, such that $\Phi^*_{\mathcal{S}}\Phi_{\mathcal{S}} = r^2$, $r\in\R$. Consequently, 
\begin{equation*}
\ee^{-\ii \lambda V_x\Phi^*_x\Phi_x} \mapsto \ee^{-\ii \lambda V_xr^2},	
\end{equation*}
and the expectation of the right-hand side exists for any potential. 

On the other hand, one can perform integration by parts to get rid of the extra factor $V_x$, since we have
\[
\ee^{-\ii \lambda V_x |\phi_x|^{2}}(1-\ii \lambda V_x \ov{\xi}_x\xi_x) = \Big(1+ \ov{\xi}_x\xi_x \frac{\partial_{\phi_x}}{\ov{\phi}_x}\Big)  \ee^{-\ii\lambda V_x |\phi_x|^{2}}.
\]
	
Integrating by parts with some care, which is needed to handle the apparently diverging contribution coming from $1/\ov{\phi}$, gives the following alternative supersymmetric representation for general random Schr\"odinger operators.
\begin{theorem}[Integration-by-parts SUSY representation]\label{thm:SUSY_formula_averaged_Greens_function}
Let $V= (V_{x})_{x \in \Z^{d}}$ be a random potential with finite-volume characteristic function
\begin{equation}\label{eq:nu-lambda}
\nu_\Lambda(r) := \E \Big[ \ee^{-\ii \sum_{x \in \Lambda} V_{x}r_{x}} \Big]\;, \qquad r= (r_{x})_{x \in \Lambda} \in \R^{\Lambda}  \;, 
\end{equation}
and let $H_0$ be a real bounded self-adjoint operator on $\ell^2(\Z^d)$. 
Then, for every finite $\Lambda\subset\Z^d$ and $\x, \y \in \Lambda$ the averaged Green's function $\mathbb E \big[G_{\Lambda}\big]$ of $H_\Lambda=(H_0+V)_\Lambda$ admits the representation for $z \in \mathbb{C}$, $\mathrm{Im} \, z >0$
\begin{equation*}\label{align:moment-free_SUSY_repr}
		\E[(G_\Lambda(z))_{\x\y}]
        \nonumber
        = \frac{\ii}{\lambda}
        \int \dPhi\Lambda
        \nu_{\Lambda}(\lvert\phi\rvert^2)\ee^{\frac{\ii}{\lambda}\phi^{*} (z-\operatorname{diag} H_0) \phi}
        \Do_{\Lambda}
        \left(\ee^{\frac{\ii}{\lambda}\Phi^*(\operatorname{diag} H_0-H_0)\Phi}\right)    
        \xi_{\x} \ov{\xi}_{\y},
\end{equation*}
where $|\phi|^{2} \equiv (|\phi_x|^{2})_{x \in \Lambda}$ and
    \[
    \Do_{\Lambda} = \prod_{x\in\Lambda} \Big(1-\ov{\xi}_x\xi_x\frac{\partial_{\phi_x}}{\ov{\phi}_x}\Big) \;.
    \]
\end{theorem}
\begin{remark}
\hspace{1em}

\begin{itemize}
\item[i)]
A key feature of Theorem~\ref{thm:SUSY_formula_averaged_Greens_function} is that the differential operator $\Do_{\Lambda}$ does not act on the diagonal contribution
\begin{equation*}
\ee^{\frac{\ii}{\lambda}\phi^{*} (z-\operatorname{diag} H_0) \phi} \;.
\end{equation*}
As a consequence, the representation is not only free of moment assumptions on the disorder distribution, but also well-suited for deriving estimates that are uniform in $E = \mathrm{Re} \, z$. 

\item[ii)] In an analogous way, one also obtains a supersymmetric representation for the averaged square modulus of the Green's function, or the average of any product of Green's funtions.
This representation also extends to models with internal degrees of freedom, with $\Do_\Lambda$ replaced by a corresponding expansion in higher-order derivatives.

\end{itemize}
\end{remark}

\paragraph{Structure of the paper}
In Section~\ref{sec:regul-DOS}, we provide a general result on the DOS for random operators, Theorem~\ref{cor:bounds}, based on suitable integrability conditions for the characteristic function of the random potential. 
We also prove that these integrability conditions are satisfied under the setting of Theorem \ref{thm:main}. 
The proof of Theorem~\ref{cor:bounds} is based on estimates on the Green's function proven in Section \ref{section:moment_free_SUSY}.
There, we prove the supersymmetric formula  of Theorem~\ref{thm:SUSY_formula_averaged_Greens_function} and use it to construct a cluster expansion that allows us to bound the Green's function. 
In Appendix~\ref{appendix:GA} we provide a short introduction to Grassmann algebras and the supersymmetric formalism.
	
\section{Regularity of the density of states}
\label{sec:regul-DOS}

\subsection{Consequences of the SUSY formalism}
\label{sec:SUSY-theorems}

In this section, we develop a criterion for the regularity of the density of states based on the supersymmetric representation in Theorem \ref{thm:SUSY_formula_averaged_Greens_function}. The assumptions are formulated directly in terms of finite-dimensional characteristic functions of the random potential and are compatible with heavy-tail distributions. 
In Section~\ref{section:application}, we verify these assumptions for the linearly correlated models considered in Theorem~\ref{thm:main}, obtaining our main result as a consequence.

The criterion is formulated in terms of the following notion.
\begin{definition}[Integrability bounds]\label{def:integrable_multiplicative_bounds}
Let $V = (V_{x})_{x \in \Z^{d}}$ be random variables and denote by $\nu_{\Lambda}$ the finite-volume characteristic functions of $V$, see \eqref{eq:nu-lambda}.
We say that $V$ satisfies integrability bounds up to order $n\in \N$ if there is a constant $K_{n}$ such that, for all finite $\Lambda \subset \Z^{d}$ and all multi-indices $p = (p_{x})_{x\in\Lambda} \in \{0,\dots,n \}^{\Lambda}$, the following bounds hold true
\begin{equation}\label{eq:integrability-bounds}
\int_{\R_{+}^{\Lambda}} \dd r \big| \nu_{\Lambda}( r^{2})\big|  \prod_{x \in \Lambda} r_{x}^{p_{x}} \leq  K_{n}^{|\Lambda|} \;,
\end{equation}
where we used the shorthand notation $r^{2} \equiv (r^{2}_{x})_{x\in\Lambda} \in \R_{+}^{\Lambda}$. We say that $V$ satisfies integrability bounds of all orders if it satisfies
integrability bounds up to order $n$, $\forall n\in\N$.
\end{definition}
\begin{remark}
Note that the argument of $\nu_{\Lambda}$ in \eqref{eq:integrability-bounds} has definite sign. If the argument were allowed to be in $\R^{\Lambda}$, this would interfere with the mixed-sign structure of the kernel $T$. 
\end{remark}
Our abstract regularity criterion is as follows.
\begin{theorem}\label{cor:bounds}
Let $H = H_{0} + \lambda V$ be an ergodic random Schr\"odinger operator on $\ell^{2}(\Z^{d})$. Assume that $H_{0}$ is a finite-range translation-invariant real self-adjoint operator and, in particular, that
\begin{equation}\label{eq:ass-H-A}
\sup_{x \in \Z^{d}} \lvert \big\{y\in\Z^d \setminus\{ x\}: (H_0)_{xy}\neq 0\big\}\rvert \leq A \;.
\end{equation}
Moreover, assume that $V$ satisfies integrability bounds up to order $2(n+A)+1$, for some $n \in \mathbb{N}$. Then, there exists $\lambda_{0}>1$ such that for every $\lambda \geq \lambda _{0} $ the DOS measure of $H$ is absolutely continuous with density $\rho \in C^{n-1}(\R)$. Moreover, Wegner-type bounds hold
        \[
        \sup_{E\in \R}\left\lvert\partial^{k}_{E}\rho(E)\right\rvert\leq \frac{c_{k}}{\lambda^{k+1}} \;, \qquad \forall 0\leq k\leq n -1 \;,
        \]
for suitable constants $c_k>1$.
\end{theorem}
The proof follows the standard strategy of deriving uniform estimates for finite-volume Green's functions and then passing to the infinite-volume limit. The key technical ingredient is the following proposition.
\begin{proposition}\label{prop:uniform_green_bounds}
Under the assumptions of Theorem \ref{cor:bounds}, there exist $\lambda_{0}>1$ and some constants $ c_k=c_k>1$ such that for every $\lambda \geq \lambda _{0} $ it holds
    \[
    \sup_{0 <\varepsilon \leq 1 }\left\lvert 
    \partial^{k}_{E}
    \E\big[(G_\Lambda( E+ \ii\varepsilon))_{ \x \x} \big]
    \right\rvert
    \leq 
    \frac{c_{k}}{\lambda^{k+1}}
    \]
for all $E\in\R$, $0 \leq k\leq n$, and $\x\in\Lambda$, uniformly in $\Lambda \subset \mathbb{Z}^{d}$.
\end{proposition}
\begin{remark}
\hspace{1em}

\begin{itemize}
\item[i)] The finite-range assumption on $H_0$ is imposed for simplicity. We expect that the result extends to short-range operators with exponentially decaying kernels. However, this extension would require additional estimates in the cluster expansion of
Section~\ref{sec:activity-estimates}, which are beyond the scope of this work.

\item[ii)] With the SUSY representation of Section \ref{section:SUSY_CE}, one can actually prove that the limit $\lim_{\veps \to 0^{+}}\partial^{k}_{E}
    \E\big[(G_\Lambda( E+ \ii\varepsilon))_{ \x \x} \big]$ exists by dominated convergence and the integrability bounds of Definition \ref{def:integrable_multiplicative_bounds}. With additional non-trivial work on the cluster expansion of Section \ref{section:SUSY_CE}, one can additionally show that also the infinite-volume limit of the previous quantities exist, see also \cite{Sjostrand1999-1}.
\end{itemize}

\end{remark}
Following standard arguments, see, e.g., \cite{klein86-,aizenman2015random}, we first show how Proposition \ref{prop:uniform_green_bounds} implies Theorem \ref{cor:bounds}. The proof of Proposition \ref{prop:uniform_green_bounds} heavily relies on the SUSY representation and SUSY cluster expansions and is thus postponed to Section \ref{section:moment_free_SUSY}.
\begin{proof}[Proof of Theorem \ref{cor:bounds}]
First of all, we have 
\begin{equation}\label{eq:observ}
 \frac{1}{|\Lambda|} \tr \E\big[G_\Lambda( E+ \ii\varepsilon) \big] = - \overline{m}_{\Lambda}(E+\ii \veps) \;,
\end{equation}
where $\overline{m}_{\Lambda}(z)$ is the Borel--Stieltjes transform associated with the averaged empirical measure $\overline{\mu}_{\Lambda}:= \mathbb{E} \, \big[\mu_{\Lambda}\big]$, compare with \eqref{eq:m-overline}. By ergodicity $(\overline{\mu}_{\Lambda})_{\Lambda}$ converges weakly to $\mu$, hence $\overline{m}_{\Lambda}(z)$ 
converges pointwise for $z \in \C^{+}$ to $m(z)$, which is in particular Herglotz. Since by Proposition \ref{prop:uniform_green_bounds} and by \eqref{eq:observ} the functions $\overline{m}_{\Lambda}$ are uniformly bounded on the upper-half plane, we have the uniform bound of the limit
\begin{equation*}
|m(E+\ii 0^{+})| \lesssim \lambda^{-1} \;, \qquad \text{for a.e.} \;E \in \R \;.
\end{equation*}
Thus, absolute continuity and boundedness of the density $\rho$ follows from the Stieltjes inversion theorem, see \cite[Appendix B]{aizenman2015random}.

Regarding the derivatives, the uniform bounds of Proposition \ref{prop:uniform_green_bounds} imply the same bounds for $\partial_{z}^{k}m(z)$, uniformly in $z \in \C^{+}$. By Fatou's theorem\footnote{See, e.g., \cite[Theorem 11.32]{Rudin} for the boundary value of an analytic function on the unit disc, which is holomorphic to the upper half plane.} we thus have convergence of $(\partial_{E}^{k}m)(E + \ii \veps)$ as $\veps \to 0^{+}$, for almost every $E \in \R$, with bounds 
\begin{equation*}
\Big|(\partial_{E}^{k}m)(E+\ii 0^{+})  \Big| \leq c_{k} \lambda^{-k-1} \;,\qquad \forall k \leq n \;.
\end{equation*}
On the other hand, for any real-valued test function $\varphi$
\begin{equation*}
\int \dd E \, (\partial_{E}^{k}m)(E+\ii \veps) \varphi(E) = (-1)^{k} \int \dd E \, m(E+\ii \veps) (\partial^{k}_{E}\varphi)(E) \;.
\end{equation*}
Taking the imaginary part, as $\veps \to 0$ we have convergence of both sides and identification of the right-hand side with integration against $\rho$ and the left-hand side as integration against the weak $k$-th derivative of $\rho$. In particular, we have that $\rho \in W^{k,\infty}(\R)$ and thus $\rho \in C^{k-1}(\R)$.
\end{proof}
To prove our main theorem, Theorem \ref{thm:main}, it remains to verify the integrability bounds for the correlated random potentials in \eqref{eq:random-V-theorem}. This is the purpose of the next section.

\subsection{Integrability bounds for linearly correlated disorder}
\label{section:application}

We now prove the required integrability bounds for the finite-volume characteristic functions of $V$.
We first verify that $V$ is a well-defined random field, which is non-trivial since $T$ is not assumed to have finite range.
\begin{lemma}\label{lemma:exist-charact}
Let $T = (T_{x y})_{x,y \in \mathbb{Z}^{d}}$ be a symmetric and periodic kernel.
Let $(W_y)_{y\in\Z^d}$ be centred i.i.d.~random variables with  characteristic function $\nu_{0}$, see \eqref{eq:nu-0-def}. Fix $\a \in (0,2]$ and assume that
\begin{equation}\label{eq:beta-stable-tail-proof}
|1-\nu_{0}(t)| \lesssim |t|^{\a}\;, \qquad \qquad \sup_{x \in \Z^{d}} \| T_{x \cdot}\|_{\Z^{d},\alpha} < \infty  \;.
\end{equation}
Then, for any $x \in \Lambda$ the random variables $V_{x}:= \sum_{ y \in \mathbb{Z}^{d}} T_{x y} \; W_{y}$ are well-defined and with characteristic function
\begin{equation*}
\nu_\Lambda(r)=\E \big[\ee ^{-\ii \sum_{x\in\Lambda}r_{x} V_{x} }\big] 
= \prod_{y \in \Z^d} \nu_0\Big(\sum_{x\in\Lambda}r_{x}T_{xy}\Big),
\qquad \forall r\in\mathbb R^\Lambda \;,
\end{equation*}
where the infinite product is understood as the limit of finite products.
\end{lemma}
\begin{proof}
We prove existence of $V_{x}$ for every $x\in \Lambda$ via L\'evy's continuity theorem applied to the sequence $(V_{x}^{(\Lambda')})_{\Lambda'}$, with $V_{x}^{(\Lambda')}:=\sum_{y \in \Lambda'}T_{xy}W_{y}$ and with $\Lambda' \nearrow \Z^{d}$. We are thus interested in proving that the truncated finite-volume characteristic functions
\begin{equation*}
\nu_{\Lambda,\Lambda'}(r) := \prod_{y \in \Lambda'} \nu_0\Big(\sum_{x\in\Lambda}r_{x}T_{xy}\Big) 
\end{equation*}
converge as $\Lambda' \nearrow \Z^{d}$ for all $r \in \R^{\Lambda}$ and that the limit is a continuous function in a neighbourhood of $r=0$. First of all, for fixed $r\in \R^{\Lambda}$, we consider $\Lambda' \supset \Lambda_{r}$, with $\Lambda_{r}$ possibly depending on $r$, and note that for some constant $C_{\Lambda,\a} \geq 1$
\begin{equation}\label{eq:first-bound}
\sum_{y \in \Lambda'\setminus \Lambda_{r}}
\Big| \sum_{x\in\Lambda}r_{x}T_{xy}\Big|^{\a}
\leq C_{\Lambda,\alpha} \Big( \sup_{x \in \Lambda} \sum_{y \in \Lambda'\setminus \Lambda_{r}} |T_{xy}|^{\a}\Big) \sum_{x\in\Lambda} |r_{x}|^{\a} \;.
\end{equation}
Then, convergence for all $r \in \R^{\Lambda}$ follows from \eqref{eq:first-bound} by choosing $\Lambda_{r} \supset \Lambda$ large enough, depending on $r$, so that $\sup_{y \in \Lambda'\setminus \Lambda_{r}}\sum_{x\in\Lambda} \big| r_{x}T_{xy}\big| \lesssim 1$, and \eqref{eq:beta-stable-tail-proof} via the following bounds
\begin{equation*}
\begin{split}
\big| \ln \nu_{\Lambda,\Lambda'}(r) - \ln \nu_{\Lambda,\Lambda_{r}}(r) \big| 
& \leq \sum_{y \in \Lambda' \setminus \Lambda_{r}} \Big| 1- \nu_0\Big(\sum_{x\in\Lambda}r_{x}T_{xy} \Big)\Big|
\\
& \lesssim \sum_{y \in \Lambda' \setminus \Lambda_{r}} \Big| \sum_{x\in\Lambda}r_{x}T_{xy} \Big|^{\a}
 \leq   C_{\Lambda,\alpha} \Big( \sup_{x \in \Lambda} \sum_{y \in \Lambda'\setminus \Lambda_{r}} |T_{xy}|^{\a}\Big) \sum_{x\in\Lambda} |r_{x}|^{\a}
\end{split}
\end{equation*}
where the r.h.s. is arbitrarily small for $\Lambda_{r}$ large enough depending on $r$.

To prove continuity at the origin, we show that convergence is uniform in a neighbourhood of $0$. By \eqref{eq:beta-stable-tail-proof}, for $r$ sufficiently small, we have
\begin{equation*}
\begin{split}
\sum_{y \in \Lambda '}\big| \ln \nu_0\Big(\sum_{x\in\Lambda}r_{x}T_{xy}\Big) \big| 
 \lesssim 
\sum_{y \in \Lambda '}\Big| \sum_{x\in\Lambda}r_{x}T_{xy}\Big|^{\a} \leq C_{\Lambda,\alpha} \big( \sup_{x \in \Z^{d}} \| T_{x \cdot}\|_{\Z^{d},\alpha}\big)^{\a} \sum_{x \in \Lambda} |r_{x}|^{\a} \;,
\end{split}
\end{equation*}
where in the last step we used the bound \eqref{eq:first-bound} with $\Lambda_{r} = \emptyset$. 
The latter bound implies convergence of the infinite product, locally uniformly, and thus continuity of this limit at the origin as $\nu_{0}$ is continuous at the origin. This concludes the proof.
\end{proof}
This also establishes the first part of Theorem \ref{thm:main}. The proof of the remaining claim of Theorem \ref{thm:main} follows from Theorem \ref{cor:bounds} and the integrability bounds proven below.
\begin{proposition}[Integrability bounds for linearly correlated potentials] \label{prop:IMB}
Under the setting of Lemma \ref{lemma:exist-charact} assume furthermore that
\begin{equation}\label{eq:beta-stable-tail-2}
|\nu_{0}(t)| \leq \ee^{-|t|^{\a}} \;,
\end{equation}
and that either, with $\kappa_{\a}=\mathbf{1}_{0 \leq \a \leq 1} + \mathbf{1}_{\a \geq 1} 2^{\frac{1}{\a}-1}$,
\begin{equation}\label{eq:condition-T}
\begin{split}
& \inf_{Y \subset \Z^{d}} \Big(
 \inf_{x \in Y} \kappa_{\a}\| T^{+}_{x \cdot} \|_{Y,1 \vee \a} - \sup_{x \in Y}\| T^{-}_{x \cdot}\|_{Y,1 \wedge \a}\Big) >0 \;, 
\\
 \mathrm{or} \qquad
& \inf_{Y \subset \Z^{d}} \Big(\inf_{x \in Y} \kappa_{\a}\| T^{-}_{x \cdot}\|_{Y,1 \vee \a} - \sup_{x \in Y}\| T^{+}_{x \cdot}\|_{Y,1 \wedge \a}\Big) >0  \;.
\end{split}
\end{equation}
Then, $V$ satisfies integrability bounds of all orders.
\end{proposition}
\begin{proof}
For any subset $Y \subset \mathbb{Z}^{d}$ and any $r \in \R_{+}^{Y}$, we will prove bounds of the form
\begin{equation}\label{eq:r-beta-bound}
\big| \nu_{Y}(r) \big| \leq \exp\big( - c \sum_{x \in Y} r_{x}^{\a} \big)\;,
\end{equation}
for some constant $c=c(T,\alpha)>0$ independent of $Y$ and $r$. Such a bound implies integrability bounds to all orders, since in fact
\[
\int_{\R_+}\d{x}\,  \ee^{-c x^{2\a}} x^n = \frac{1}{2\a c^{(n+1)/2\a}}\Gamma \Big(\frac{n+1}{2\a}\Big) \;, \qquad \forall n \in \N \;,
\]
where $\Gamma$ is the Euler Gamma function.
\medskip

To prove bound \eqref{eq:r-beta-bound}, by \eqref{eq:beta-stable-tail-2} we have
\begin{equation*}
\big| \nu_{Y}(r) \big| =
\prod_{z \in \Z^{d}} \big| \nu_{0}(\sum_{x \in Y} r_{x} T_{x z}) \big| \leq \exp \Big( -\sum_{z \in \Z^{d}} \big| \sum_{x \in Y} r_{x} T_{x z} \big|^{\a} \Big) 
\leq 
\exp \Big( -\sum_{z \in Y} \big| \sum_{x \in Y} r_{x} T_{x z} \big|^{\a} \Big) 
 \;.
\end{equation*}
Therefore, our goal is to show that 
\begin{equation}\label{eq:main-proof-a-small}
\sum_{z \in Y} \big| \sum_{x \in Y} r_{x} T_{x z} \big|^{\a} \geq c \sum_{x \in Y} r_{x}^{\a} \;,
\end{equation}
where $c = c(T,\a)$ is independent of $Y$ and $r$. We will work under the assumption that the first condition in \eqref{eq:condition-T} is satisfied.
The case with $T^{+}$ and $T^{-}$ swapped follows by replacing $T$ with $-T$ in \eqref{eq:main-proof-a-small}.
We distinguish the cases $0<\alpha \leq 1$ and $1< \alpha \leq 2$.
\medskip

\noindent
\underline{Case $0 < \alpha \leq 1$.} 
The proof of \eqref{eq:main-proof-a-small} is a based on concavity/subadditivity of the map $t \mapsto t^{\a}$. Subadditivity implies the following bound for any nonnegative $(b_{x})_{x}$ 
\begin{equation}\label{eq:sub}
\begin{split}
\sum_{x} b_{x}^{\a} & \geq \Big( \sum_{x} b_{x} \Big)^{\a} \;,
\end{split}
\end{equation}
whereas concave Jensen's inequality gives, for non-negative $(\theta_{x})_{x}$ such that $\sum_{x} \theta_{x} =1$
\begin{equation}\label{eq:con}
\begin{split}
\Big( \sum_{x} \theta_{x} b_{x} \Big)^{\a} & \geq \sum_{x} \theta_{x} b_{x}^{\a}  \;.
\end{split}
\end{equation}
First of all, we use subadditivity to split the contributions of $T^{+}$ and $T^{-}$,
\begin{equation*}
\Big| \sum_{x \in Y} r_{x}T_{xz} \Big|^{\a} 
\geq 
 \Big(\sum_{x \in Y } r_{x} T^+_{xz}\Big)^{\a} - \Big(\sum_{x \in Y } r_{x} T^-_{xz}\Big)^{\a} \;.
\end{equation*}
We control the contribution due to $T^{-}$ by \eqref{eq:sub}
\begin{equation*}
\begin{split}
\sum_{z \in Y} \Big(\sum_{x \in Y } r_{x} T^-_{xz}\Big)^{\a} &\leq \sum_{z \in Y}  \sum_{x \in Y } r_{x}^{\a} (T^-_{xz})^{\a}
 =  \sum_{x \in Y } r_{x}^{\a} \| T^{-}_{x \cdot} \|^{\alpha}_{Y,1 \wedge \alpha}
 \leq \Big( \sup_{x \in Y } \| T^{-}_{x \cdot} \|_{Y,1 \wedge \alpha}\Big)^{\alpha} \sum_{x \in Y } r_{x}^{\a} \;.
\end{split}
\end{equation*}
On the other hand, we control the contribution due to $T^{+}$ by exploiting its entrywise non-negativity and \eqref{eq:con}
\begin{equation*}
 \Big( \sum_{x \in Y} r_{x} \frac{T^{+}_{xz}}{\sum_{x \in Y} T^{+}_{xz}}\Big)^{\a} \geq \sum_{x \in Y} r_{x}^{\a} \frac{T^{+}_{xz}}{\sum_{x \in Y} T^{+}_{xz}}  \;,
\end{equation*}
or, equivalently, using furthermore that $T^{+}$ is symmetric
\begin{equation*}
\begin{split}
\Big( \sum_{x \in Y} r_{x}T^{+}_{xz}\Big)^{\a} 
& \geq \|T^{+}_{z \cdot} \|_{Y,1}^{\a -1} \sum_{x \in Y} r_{x}^{\a} T^{+}_{xz} \;.
\end{split}
\end{equation*}
Thus, 
\begin{equation*}
\begin{split}
\sum_{z \in Y}\Big( \sum_{x \in Y} r_{x}T^{+}_{xz}\Big)^{\a} 
& \geq 
\Big(\inf _{z \in Y}
\|T^{+}_{z \cdot} \|_{Y,1}^{\a -1}  \Big)\sum_{z \in Y} \sum_{x \in Y} r_{x}^{\a} T^{+}_{xz} 
\\
&  = \Big( \inf _{z \in Y} \|T^{+}_{z \cdot } \|_{Y,1 \lor \alpha} \Big)^{\a -1} \sum_{x \in Y} r_{x}^{\a}  \|T^{+}_{x \cdot} \|_{Y,1 \lor \alpha} 
 \geq
\Big( \inf _{z \in Y} \|T^{+}_{z \cdot} \|_{Y,1 \lor \alpha} \Big)^{\a} \sum_{x \in Y} r_{x}^{\a} \;.
\end{split}
\end{equation*}
All in all, using monotonicity of , we obtain \eqref{eq:main-proof-a-small} with
\begin{equation*}\label{eq:partial-neg-beta}
\begin{split}
c = \inf_{Y} \bigg[
\Big( \inf _{x \in Y} \|T^{+}_{x \cdot} \|_{Y,1 \lor \alpha} \Big)^{\a}
- \Big( \sup_{x \in Y } \| T^{-}_{x \cdot} \|_{Y,1 \land \alpha}\Big)^{\alpha} \bigg] \;,
\end{split}
\end{equation*}
which is positive by assumptions on $T$, see \eqref{eq:condition-T}.
\medskip

\noindent
\underline{Case $1 < \alpha \leq 2$.} The proof is a based on convexity/superadditivity of the map $t \mapsto t^{\a}$. Superadditivity implies the following bound for any nonnegative $(b_{x})_{x}$ 
\begin{equation}\label{eq:super}
\begin{split}
\sum_{x} b_{x}^{\a} & \leq \Big( \sum_{x} b_{x} \Big)^{\a} \;,
\end{split}
\end{equation}
whereas convex Jensen's inequality gives, for non-negative $(\theta_{x})_{x}$ such that $\sum_{x} \theta_{x} =1$
\begin{equation}\label{eq:con-con}
\begin{split}
\Big( \sum_{x} \theta_{x} b_{x} \Big)^{\a} & \leq \sum_{x} \theta_{x} b_{x}^{\a}  \;.
\end{split}
\end{equation}
We proceed as before, separating the positive and negative contributions of $T$
\begin{equation*}
\Big| \sum_{x \in Y} r_{x}T_{xz} \Big|^{\a} 
\geq 
2^{1-\a}\Big(\sum_{x \in Y } r_{x} T^+_{xz}\Big)^{\a} - \Big(\sum_{x \in Y } r_{x} T^-_{xz}\Big)^{\a} \;,
\end{equation*}
where this time we paid an additional factor because for $\alpha \geq 1$ genuine subadditivity does not hold.
We control the contribution due to $T^{+}$ by \eqref{eq:super} 
\begin{equation*}
\sum_{z \in Y}\Big(\sum_{x \in Y } r_{x} T^+_{xz}\Big)^{\a} \geq \sum_{z \in Y} \sum_{x \in Y } r_{x}^{\a} (T^+_{xz})^{\a} \geq
\Big( \inf _{x \in Y} \|T^{+}_{x \cdot} \|_{Y,1 \lor \alpha} \Big)^{\a} \sum_{x \in Y} r_{x}^{\a} \;,
\end{equation*}
where in the last step we also used that $T^{+}$ is a symmetric kernel.
We control the contribution due to $T^{-}$ by \eqref{eq:con-con}. Proceeding as before, we find
\begin{equation*}
\sum_{z \in Y} \Big(\sum_{x \in Y } r_{x} T^{-}_{xz}\Big)^{\a} 
\leq  \Big(\sup_{x \in Y} \|T^{-}_{x \cdot} \|_{Y, 1 \land \a } \Big)^{\a} \sum_{z \in Y} r_{x}^{\a} \;,
\end{equation*}
where we also used that $T^{-}$ is a symmetric kernel. All in all, we have
\begin{equation*}
\sum_{z \in \Z^{d}}\Big| \sum_{x \in Y} r_{x}T_{xz} \Big|^{\a} 
 \geq  \bigg [ \Big(2^{\frac{1}{\a}-1}  \inf _{x \in Y} \|T^{+}_{x \cdot} \|_{Y,1 \lor \alpha} \Big)^{\a}
- \Big( \sup_{x \in Y } \| T^{-}_{x \cdot} \|_{Y,1 \land \alpha}\Big)^{\alpha} \bigg] \sum_{x\in Y}  r_x^\a \;,
\end{equation*}
implying \eqref{eq:r-beta-bound} as before. This concludes the proof.
\end{proof}
\begin{remark}\label{eq: block-case}
By the triangle inequality, we note the crude bound
\begin{equation}\label{eq:naive}
\sum_{z \in Y} \Big \|\sum_{x \in Y} r_{x}T_{x z} \Big\| \geq  \Big \|\sum_{x \in Y} r_{x} \sum_{z \in Y} T_{x z} \Big\| \;,
\end{equation}
where we used the notation $\| a\| := \mathrm{d}(0,a)$ for some metric $\mathrm{d}(\cdot , \cdot)$ on $\R$.  An interesting case of metric would be $\mathrm{d}(a,b):= \ell \log(1+|a-b|)$ for some $\ell \in \N$. In fact, if $|\nu_{0}(t)| \leq (1+|t|)^{-\ell}$ and $\sum_{z \in Y} T_{x z} = c \neq 0$, \eqref{eq:naive} can be used to derive the estimate for $r \in \R_{+}^{Y}$
\begin{equation*}
\big| \nu_{Y}(r) \big| \leq \exp \Big( - \sum_{z \in Y} \Big \|\sum_{x \in Y} r_{x}T_{x z} \Big\| \Big) \leq \Big(\frac{1}{1+ c \sum_{x \in Y} r_{x} }\Big)^{\ell} \lesssim_{Y} \prod_{x \in Y} \frac{1}{1+ c r_{x}^{\ell/|Y|}} \;.
\end{equation*}
This is useless in the general case, as $Y$ can be arbitrarily large. However, if $T$ is block diagonal with blocks of size $Q$, $\nu_{Y}$ factorises over the blocks, so that the above inequality gives integrability bounds up to a certain power for $\ell/Q$ large enough.
\end{remark}
	
\section{SUSY approach for heavy-tailed disorder}
\label{section:moment_free_SUSY}

\subsection{SUSY representation and cluster expansion}
\label{section:SUSY_CE}

In this section, we prove Theorem~\ref{thm:SUSY_formula_averaged_Greens_function} and provide a cluster-expansion formula for the averaged Green's function that can be applied to heavy-tailed disorder. 
Throughout the rest of the section, we work with the supersymmetric formalism; we refer to Appendix \ref{appendix:GA} for details.
  
\begin{proof}[Proof of Theorem~\ref{thm:SUSY_formula_averaged_Greens_function}]
We write the Green's function as
     \[
     G_\Lambda(E+ \ii\varepsilon) = \frac{\ii}{\lambda}\left(\frac{\varepsilon + \ii(H_0)_\Lambda-\ii E}{\lambda}+\ii V\right)^{-1} \;,
     \]
so that, by Lemma~\ref{lemma:SUSY_Gauß_int} we obtain the integral representation, for any $\x, \y \in \Lambda$
\begin{equation}\label{eq:SUSY_repr_Greens_function}
        \big(G_\Lambda(E+\ii \varepsilon) \big)_{\x\y}
		= \frac{\ii}{\lambda}\int \dPhi{\La} \ee^{\frac{\ii}{\lambda}\Phi^*(E+\ii\varepsilon-H_0-\lambda V)\Phi} \xi_{\x}\ov{\xi}_{\y}  \;.
\end{equation}
Next, we want to average the expression (\ref{eq:SUSY_repr_Greens_function}). As discussed in Section \ref{sec:main-result}, one can only pull the expectation into the integral if $V$ admits a sufficient number of moments. To circumvent this problem, we perform integration by parts. First of all, we abridge
\begin{equation}\label{eq:def-P-zeta}
\zeta := E+\ii\varepsilon-\operatorname{diag} H_0 \;,
\qquad
P:=\operatorname{diag}H_0-H_0 \;.
\end{equation}
so that, expanding the exponential we have
\begin{equation}\label{eq:proof-representation-parts}
\begin{split}
 \ee^{\frac{\ii}{\lambda}\Phi^*(E+\ii\varepsilon-H_0-\lambda V)\Phi}
&=
\left(\prod_{x\in\Lambda}\Big(1+\frac{\ii}{\lambda} (\zeta_{xx}-\lambda V_x)\ov{\xi}_x\xi_x\Big)\ee^{\frac{\ii}{\lambda} (\zeta_{xx}-\lambda V_x)\ov{\phi}_x\phi_x} \right) \ee^{\frac{\ii}{\lambda}\Phi^*P\Phi} 
\\
& = \left(\prod_{x\in\Lambda}\Big(1+\ov{\xi}_x\xi_x\frac{\partial_{\phi_x}}{\ov{\phi}_x}\Big)\ee^{\frac{\ii}{\lambda} (\zeta_{xx}-\lambda V_x)\ov{\phi}_x\phi_x}\right) \ee^{\frac{\ii}{\lambda}\Phi^*P\Phi}
\\
& =\Do_{\Lambda}^{*} (\ee^{\frac{\ii}{\lambda}\phi^*(\zeta-\lambda V)\phi})\, \ee^{\frac{\ii}{\lambda}\Phi^*P\Phi}
\end{split}
\end{equation}
where
\[
 \Do_{\Lambda}^{*} = \prod_{x\in\Lambda}\Big(1+\ov{\xi}_x\xi_x\frac{\partial_{\phi_x}}{\ov{\phi}_x}\Big).
\]
Therefore, noting that $1/\ov{\phi}_x$ does not lead to divergences because of integration with respect to $\dd \phi_{x} = \frac{\dd \mathrm{Re} \phi_{x} \dd \mathrm{Im} \phi_{x}}{\pi}$ and performing integration by parts, we obtain
\begin{equation}\label{eq:Green_with_adjoint}
\int \dPhi{\La} \ee^{\frac{\ii}{\lambda}\Phi^*(E+\ii\varepsilon-H_0-\lambda V)\Phi}\xi_{\x}\ov{\xi}_{\y}
        	= 
        	\int \dPhi{\Lambda} \ee^{\frac{\ii}{\lambda}\phi^*(\zeta-\lambda V)\phi} \Do_{\Lambda}(\ee^{\frac{\ii}{\lambda}\Phi^*P\Phi}) \xi_{\x}\ov{\xi}_{\y} \;,
\end{equation}
where
\[
        \Do_{\Lambda} = \prod_{x\in\Lambda}\Big(1-\ov{\xi}_x\xi_x\frac{\partial_{\phi_x}}{\ov{\phi}_x}\Big),
\]
is the adjoint of $\Do_{\Lambda}^{*}$, and where the boundary term is vanishing because $(\Re \phi, \Im \phi) \to \ee^{\ii\phi^*\zeta\phi}$ is a Schwartz function. 

We conclude by taking the expectation inside of the r.h.s.~of (\ref{eq:Green_with_adjoint}) by Fubini. This is justified because $\lVert \Do_{\Lambda}^*\ee^{\frac{\ii}{\lambda}\Phi^*P\Phi}\rVert$ grows at most polynomially in $\phi$ (with order possibly depending on $\Lambda$), see Appendix \ref{appendix:GA} for the definition of the norm, and $|\ee^{\frac{\ii}{\lambda}\phi^*(\zeta-\lambda V)\phi}| \leq \prod_{x \in \Lambda} \ee^{-\frac{\varepsilon}{\lambda} |\phi_{x}|^{2}}$, so that
\[
            (\Re \phi, \Im \phi) \mapsto \sup_{V} \Big|  \ee^{\frac{\ii}{\lambda}\phi^*(\zeta-\lambda V)\phi} \Big|
             \left\lVert \Do_{\Lambda}^* \ee^{\frac{\ii}{\lambda}\Phi^*P\Phi}\right\rVert \in L^1(\R^{2\LC} ) \;.
\] 
\end{proof}
The main ingredient to prove Proposition~\ref{prop:uniform_green_bounds} is a cluster expansion applied to the SUSY representation of the averaged Green's function of Theorem \ref{thm:SUSY_formula_averaged_Greens_function}.     
We employ a Battle--Brydges--Federbush cluster expansion \cite[Theorem 3.1]{Brydges1984}, which we quickly recall for the reader's convenience. For an subset $X\subset \Lambda$ let $\mathcal{V}_X := \frac{1}{2}\sum_{x,y\in X} v_{xy}$ be a two-body potential, with $v_{xy}=v_{yx}$, $v_{xx} = 0$, even elements of a Grassmann algebra. One can expand $\exp\big(\mathcal V_X\big)$ into connected components
\[
\ee^{\mathcal V_X} = \sum_{\Pi\in\operatorname{Part}(X)} \prod_{Y\in \Pi} \big(\ee^{\mathcal V_Y} \big)_{\mathrm{conn}},	
\]
where $\operatorname{Part}(X)$ denotes the partitions of $X$ and where $\big(\ee^{\mathcal V_Y} \big)_{\mathrm{conn}}$ are the connected parts, defined in a recursive way. The recursive definition is difficult to employ in applications and a more convenient representation is as follows.
\begin{theorem}[{\cite[Theorem 3.1]{Brydges1984}}]\label{thm:BBF} 
The following identity holds true
\begin{equation*}
\big(\ee^{\mathcal V_Y} \big)_{\mathrm{conn}} = \sum_{T \;\mathrm{ on }\;Y}\left(\prod_{(x,y)\in T}v_{xy}\right)\int\d{\rho_T}(s)\,\ee^{\mathcal V_Y(s)}\;.
\end{equation*}
Here, the sum runs over trees $T$ on $Y$, $\d{\rho_T}$ is a probability measure supported on $[0,1]^{P(Y)}$, where $P(Y)$ is the set of unordered pairs in $Y$, and
\[
\mathcal V_Y(s) := \frac{1}{2}\sum_{x,y\in Y} s_{xy}v_{xy}, \qquad s = (s_{xy})\in[0,1]^{P(Y)},		
\]
is a convex decoupling of $\mathcal  V_Y$, i.e., $\mathcal  V_Y(s)$ can be written as a convex linear combination of terms of the form $\sum_{Y'\in\Pi}\mathcal V_{Y'}$ where $\Pi$ is a partition of $Y$.
\end{theorem} 
This result is presented in \cite[Theorem 3.1]{Brydges1984} for complex valued two-body potential $\mathcal{V}_{X}$, but the statement really applies to commutative Banach algebras, and, in particular, to even Grassmann algebras, compare with \cite{Fresta2021}.
\begin{proposition}[Integration-by-parts SUSY cluster expansion]\label{prop:moment_free_cluster_expansion}
Under the assumptions of Theorem \ref{thm:SUSY_formula_averaged_Greens_function}, let $\zeta$ and $P$ be as in \eqref{eq:def-P-zeta} and set
\[
\mathcal{V}_{Y}:= \frac{1}{2} \sum_{x,y \in Y} v_{xy} \;,
\qquad
v_{xy} := \Phi^*_{x} P_{xy} \Phi_{y} + \Phi^*_{y} P_{yx} \Phi_{x} \;.
\]
Then, for any $\varepsilon >0$, $n \geq 0$ and $\x\in\Lambda$  we have the identity
\begin{equation}\label{eq:prop_cluster_exp_sum_over_KT(Y)}
   \partial^{n}_{E}
    \E[(G_\Lambda(E+ \ii \varepsilon))_{\x\x}]
    =
    2\sum_{\substack{Y\subset\Lambda\colon\\ Y \ni \x}}  \left(\frac{\ii}{\lambda}\right)^{\lvert Y\rvert+n}
            \sum_{T \;\mathrm{ on }\;Y} \mathcal{K}_{T,n}(Y),
\end{equation}
where
\begin{equation*}
\mathcal{K}_{T,n}(Y) =
			\int  \d{\rho_T}(s) 
             \int \dPhi{Y}
             \nu_{Y}(\lvert\phi\rvert^2) \big( \sum_{x \in Y}  |\phi_{x}|^{2}\big)^{n}
            \ee^{\frac{\ii}{\lambda}(\phi^{*} \zeta \phi)_{Y}} 
             \Do_Y\left(\prod_{(x,y)\in T}v_{xy}\, \ee^{ \frac{\ii}{\lambda}\mathcal V_Y(s)}\right)\xi_{\x}\ov{\xi}_{\x} \;.
\end{equation*}
having used the shorthand $(\phi^{*} \zeta \phi)_{Y} = \sum_{x \in Y} \zeta_{x} |\phi_{x}|^{2} $.
\end{proposition}
\begin{proof}
The proof is an application of Theorem \ref{thm:BBF} and the supersymmetric localisation theorem, see Theorem \ref{theorem:SUSY_Localization}. Our starting point is the supersymmetric representation \eqref{eq:SUSY_repr_Greens_function} before taking the disorder average. This is because disorder expectation introduces correlations and it would be harder to keep track of the needed supersymmetric cancellations.

Define the diagonal matrix $\tilde{\zeta}:= \zeta - \lambda V$ and expand the Gibbs factor $\ee^{\frac{\ii}{\lambda}  \Phi^{*}P \Phi}$ inside of \eqref{eq:SUSY_repr_Greens_function} in connected components, which we write by means of Theorem \ref{thm:BBF} 
\begin{align*}\label{eq:expected_of_pair_with_clusterexp}
            &(G_\Lambda(E+\ii\varepsilon))_{\x\x}
            \\
            &
            =
            \frac{\ii}{\lambda}
            \int \dPhi{\La} \ee^{\frac{\ii}{\lambda} \Phi^{*} \tilde{\zeta} \Phi} \xi_{\x} \ov{\xi}_{\x}
            \left[
            \sum_{\Pi\in\operatorname{Part}(\Lambda)}\prod_{Y\in\Pi}\sum_{T\text{ on }Y}
			\int \d{\rho_T}(s)
			\left(\prod_{(x,y)\in T}\frac{\ii v_{xy}}{\lambda}\right)
			\ee^{\frac{\ii}{\lambda}\mathcal{V}_{Y}(s)}
            \right]
            \\
            &=
            \frac{\ii}{\lambda}
            \sum_{\Pi\in\operatorname{Part}(\Lambda)}\prod_{Y\in\Pi}\sum_{T\text{ on }Y}
            \int \d{\rho_T}(s)
            \int \dPhi{Y} \ee^{\frac{\ii}{\lambda} (\Phi^{*} \tilde{\zeta} \Phi)_{Y}} (\xi_{\x}\ov{\xi}_{\x})^{\mathbf{1}_{\x\in Y}}
            \left(\prod_{(x,y)\in T}\frac{\ii v_{xy}}{\lambda}\right)
			\ee^{\frac{\ii}{\lambda}\mathcal{V}_{Y}(s)} \;,
\end{align*}
where we factorised integration in $\dd \Phi_{\Lambda}$, the term $\ee^{\frac{\ii}{\lambda} \Phi^{*} \tilde{\zeta} \Phi} = \prod_{Y}\ee^{\frac{\ii}{\lambda} \big(\Phi^{*} \tilde{\zeta} \Phi\big)_{Y}} $, and switched order of integration by Fubini, compare with the proof of Theorem \ref{thm:SUSY_formula_averaged_Greens_function}.
If $\x\notin Y$, then the  integrand is supersymmetric and decaying at infinity $\mathbb{P}$-almost surely in the disorder. In this case, by Theorem~\ref{theorem:SUSY_Localization}, the integral evaluates to
\begin{equation*}
		\int \dPhi{Y} \ee^{\frac{\ii}{\lambda} \big(\Phi^{*} \tilde{\zeta} \Phi\big)_{Y}}
            \left(\prod_{(x,y)\in T}\frac{\ii v_{xy}}{\lambda}\right)
			\ee^{\frac{\ii}{\lambda}\mathcal{V}_{Y}(s)}
        = \begin{cases}
            0 & \text{if } \lvert Y\rvert > 0\\
            1 & \text{if } \lvert Y\rvert = 0.
        \end{cases}
\end{equation*}
Consequently, the product over $Y\in\Pi$ is zero if there exists a set $Y\in\Pi$ such that $\x\notin Y$ and $\lvert Y\rvert >1$. In other words, the only partitions which give nonzero contributions are of the form $\Pi = \{Y, \{x_1\},\ldots,\{x_N\}\}$, where $Y\subset \Lambda$ such that $\x\in Y$ and $\{x_1,\ldots,x_N\} = \Lambda\setminus Y$. 
Finally, to obtain (\ref{eq:prop_cluster_exp_sum_over_KT(Y)}) for $n=0$, we do partial integration and take the disorder expectation inside the integral, as in the proof of theorem~\ref{thm:SUSY_formula_averaged_Greens_function}.
The identity for $n>0$ is obtained by taking the derivatives inside the integral by dominated convergence, since $\varepsilon >0$.
\end{proof}

\subsection{Activity estimates: proof of Proposition~\ref{prop:uniform_green_bounds}}
\label{sec:activity-estimates}
	
The cluster expansion of Proposition~\ref{prop:moment_free_cluster_expansion}
expresses the averaged Green's function as a sum over clusters connected by the branches of the trees $T$. The convergence of this expansion therefore follows from suitable estimates on the cluster activity $\mathcal{K}_{T,n}(Y)$. We first derive such a bound for a fixed tree $T$, keeping explicit the dependence on the hopping amplitudes and on the moments of the characteristic functions of $V$.
\begin{proposition}\label{prop:bound_on_K_T,alpha}
Under the assumptions of Theorem \ref{cor:bounds}, assume further that $\lambda \geq 1$. Then, there exists a constant $C>1$, depending only on $H_{0}$, such that, for any $Y \subset \Z^{d}$ finite, $T$ on $Y$, and $n \in \N_{0}$
\begin{equation*}
           \lvert \mathcal{K}_{T,n}(Y)\rvert 
            \leq C^{\lvert Y\rvert}\left(\prod_{(x,y)\in T} |P_{xy}| \right)
          \int_{\R_{+}^{Y}} \d{r}\, \big| \nu_{Y}(r^{2}) \big|  \big( \sum_{x \in Y}  r^{2}_{x}\big)^{n}\prod_{x\in Y}\big( 1 + r_x^{2A+1}\big)
  \;.
\end{equation*}
\end{proposition}
\begin{proof}
We denote by $C$ generic positive constants depending only on $H_{0}$, whose value may change from line to line. By Lemma \eqref{lemma:bound-norm} we have the bound
\begin{equation}\label{eqLfirst-bound-K}
\lvert \mathcal{K}_{T,n}(Y)\rvert \leq 
   	\int  \d{\rho_T}(s) 
   	\int \dphi{Y}\,
   \big|	\nu_{Y}(\lvert\phi\rvert^2) \big| \big( \sum_{x \in Y}  |\phi_{x}|^{2}\big)^{n}
   	\left\lVert\Do_Y\left(\prod_{(x,y)\in T}v_{xy}\, \ee^{\mathcal V_Y(s)}\right)\xi_{\x}\ov{\xi}_{\x}\right\rVert \;.
\end{equation}
To estimate the Grassmann norm, we introduce the shorthand notation
   	\begin{align*}
   	h_T(\Phi):=\prod_{(x,y)\in T} \left(\Phi^*_xP_{xy}\Phi_{y}+\Phi^*_{y}P_{yx}\Phi_x\right), \qquad
   	f_T(\phi) :=
   	\ee^{\frac{\ii}{2\lambda}\sum_{x,y\in Y}s_{xy}\left(\ov{\phi}_xP_{xy}\phi_{y}+\ov{\phi}_{y}P_{yx}\phi_x\right)} \;,
   	\end{align*}
and expand the differential operator 
   	\[
   	\Do_Y=\prod_{x\in Y} \Big(1-\ov{\xi}_x\xi_x\frac{\partial_{\phi_x}}{\ov{\phi}_x}\Big) = \sum_{\beta\in\{0,1\}^{Y}} (-1)^{\lvert \beta\rvert}\prod_{x\in Y}\frac{(\ov{\xi}_x\xi_x)^{\beta_x}}{\ov{\phi}_x^{\beta_{x}}} \partial_\phi^\beta \;.
   	\]
Using the sub-multiplicativity property of the norm, see \eqref{eq:product-norm}, we have
\begin{align}\label{eq:bound-2-K}
   	\left\lVert\Do_Y\left(\dots\right)\xi_{\x}\ov{\xi}_{\x}\right\rVert
   	&\leq
   	\sum_{\beta\in\{0,1\}^{Y}}
   	\frac{1}{\prod_{x\in Y}\lvert\phi_x\rvert^{\beta_x}} \left\lVert\partial_\phi^\beta\left(h_T(\Phi)f_T(\phi)\right)\right\rVert 
   	\\
   	&\hspace{3cm}
   	\times \left\lVert\prod_{x\in\Lambda}(\ov{\xi}_x\xi_x)^{\beta_x}\xi_{\x}\ov{\xi}_{\x}\ee^{\frac{\ii}{2\lambda}\sum_{x,y\in Y}s_{xy}\left(\ov{\xi}_xP_{xy}\xi_{x}+\ov{\xi}_{y}P_{yx}\xi_x\right)}\right \rVert. \nonumber
   	\end{align}
   	The purely Grassmann term is estimated by repeated application of the sub-multiplicativity property, and by using that $\|\xi_{x}\| =\| \ov{\xi}_{x}\| = 1$ for any $x$
   	\begin{equation}\label{eq:bound-K-grass}
   	\begin{split}
   	&\left\lVert 
		\prod_{x\in Y}(\ov{\xi}_x\xi_x)^{\beta_x}
		\xi_{\x}\ov{\xi}_{\x}
        \ee^{\frac{\ii}{2\lambda}\sum_{x,y\in Y}s_{xy}\left(\ov{\xi}_xP_{xy}\xi_{y}+\ov{\xi}_{y}P_{yx}\xi_x\right)}\right\rVert
        \\
& \qquad \leq \left\lVert 
        \ee^{\frac{\ii}{2\lambda}\sum_{x,y\in Y}s_{xy}\left(\ov{\xi}_xP_{xy}\xi_{y}+\ov{\xi}_{y}P_{yx}\xi_x\right)}\right\rVert
\leq
 \ee^{\frac{1}{\lambda} \sum_{x,y \in Y} |P_{xy}|} \leq C^{|Y|} \;,
\end{split}
   	\end{equation}
where we also used that $s_{xy}\leq 1$ and that for a bounded finite-range operator $\sum_{x,y \in Y} |P_{xy}| \leq C |Y| $. 
On the other hand, by the Leibniz rule for derivatives
    \begin{equation*}
    	\left\lVert\partial_\phi^\beta\left(h_T(\Phi)f_T(\phi)\right)\right\rVert
    	\leq
    	\sum_{\gamma\leq\beta} \binom{\beta}{\gamma}
    	\left\lVert \partial_\phi^\gamma h_T(\Phi) \right\rVert
    	\left\lvert 
    	\partial_\phi^{\beta-\gamma} f_T(\phi)
    	\right\rvert.
    \end{equation*} 
where for multi-indices $\gamma,\beta \in \{ 0,1\}^{Y}$ we introduced the binomial coefficient
\begin{equation*}
\binom{\beta}{\gamma}:= \prod_{x} \binom{\beta_{x}}{\gamma_{x}} = 1 \;,
\end{equation*}
and where the notation $\gamma \leq \beta$ means $\gamma_{x}\leq \beta_{x}$ for all $x \in Y$. 
An explicit computation gives
    \begin{align*}
    \lvert
    \partial_\phi^{\beta-\gamma} f_T(\phi)
    \rvert
    & =
    \left\lvert
    \prod_{x\in Y }\left(\frac{\ii}{\lambda}\sum_{y\in Y} s_{xy}\ov{\phi}_yP_{yx}\right)^{\beta_x-\gamma_x}f_T(\phi)\right\rvert
    =
    \prod_{x\in Y } \left\lvert\frac{1}{\lambda}\sum_{y\in Y} s_{xy}\ov{\phi}_yP_{yx}\right\rvert^{\beta_x-\gamma_x} \;,
    \end{align*}
so that, using the finite-range property of $H_{0}$, see \eqref{eq:ass-H-A}, we obtain
    \begin{equation*}
    \lvert
    \partial_\phi^{\beta-\gamma} f_T(\phi)
    \rvert \leq  C^{|Y|} \lambda^{-\lvert\beta-\gamma\rvert}
    \prod_{x\in Y}(\lvert \phi_x\rvert^{A}\lor 1) \leq 
     C^{|Y|} \prod_{x\in Y}(\lvert \phi_x\rvert^{A}\lor 1) 
    \end{equation*}
for some constant $C$ depending on $H_{0}$, where we used the notation $|\beta-\gamma| = \sum_{x}(\beta_{x} - \gamma_{x})$.
   
To bound the contribution due to $h_T(\Phi)$, we note that by the finite-range assumption on $H_{0}$ and hence on $P$, each vertex of the tree has at most coordination $A$. Using once more that the Grassmann variables have norm one, we have 
    \[
    \left\lVert \partial_\phi^\gamma h_T(\Phi) \right\rVert
    \leq
    C^{\lvert Y\rvert}\left(\prod_{(x,y)\in T} |P_{xy}| \right)A^{\lvert \gamma\rvert}\prod_{x\in Y}(\lvert \phi_x\rvert^{A}\lor 1).
    \]
Thus, using that $|\beta|\leq |Y|$
\begin{equation}\label{eq:K-last-step}
 \left\lVert\partial_\phi^\beta\left(h_T(\Phi)f_T(\phi)\right)\right\rVert
    \leq 
    C^{\lvert Y\rvert}
    \left(\prod_{(x,y)\in T} |P_{xy}| \right)
    \prod_{x\in Y}(\lvert \phi_x\rvert^{2A}\lor 1).
\end{equation}
for some constant depending on $H_{0}$.
    Putting together \eqref{eqLfirst-bound-K}, \eqref{eq:bound-2-K}, \eqref{eq:bound-K-grass}, and \eqref{eq:K-last-step}, and rearranging the sum over $\beta$, we obtain
    \begin{equation*}
        \begin{split}
        \lvert \mathcal{K}_{T,n}(Y)\rvert
        & \leq
        C^{\lvert Y\rvert}
        \left(\prod_{(x,y)\in T} |P_{xy}| \right)
        \int \dphi{Y}
            \big|\nu_{Y}(\lvert\phi\rvert^2)\big|
\big( \sum_{x \in Y}  |\phi_{x}|^{2}\big)^{n}            
             \prod_{x\in Y}\Big(1+\frac{1}{|\phi_{x}|}\Big)(\lvert \phi_x\rvert^{2A}\lor 1)
            \end{split}
    \end{equation*}
We write the integral in polar coordinates $r_{x} = |\phi_{x}|$, and use that $\Big(1+\frac{1}{|\phi_{x}|}\Big) (\lvert \phi_x\rvert^{2A}\lor 1) \lesssim (\lvert \phi_x\rvert^{2A} + \frac{1}{|\phi_{x}|}) $ to obtain the claim.
\end{proof}
We now prove Proposition  \ref{prop:uniform_green_bounds} using moment bounds for the characteristic functions of $V$ and standard tree-stripping estimates.
\begin{proof}[Proof of Proposition~\ref{prop:uniform_green_bounds}]
We rearrange the expansion from Proposition~\ref{prop:moment_free_cluster_expansion} by writing
\begin{align*}
   & \sum_{\substack{Y\subset\Lambda \colon\\ Y \ni \x}}\;\sum_{T\text{ on }Y} \mathcal{K}_{T,n}(Y) 
   = 
		\sum_{N = 0}^{\lvert \Lambda'\rvert} \frac{1}{N!} \sum_{\substack{y_1,\ldots,y_N\in\Lambda'\\ \text{ distinct}}}
		\;
		\sum_{T\text{ on }\{1,\ldots, N+1\}} \mathcal{K}_{T,n}(\{y_1,\ldots,y_{N+1}\}),
\end{align*}
where we have set $\Lambda' = \Lambda \setminus \{\x\}$ and $y_{N+1} = \x$. The factor $1/N!$ makes sure that every configuration is counted only once.

Thus, using also Proposition \ref{prop:bound_on_K_T,alpha}, we obtain the bound 
\begin{align}\label{eq: semifinal-G-bound}
        &\left\lvert \partial^{n}_{E}\E[(G_\Lambda(E+ \ii\varepsilon))_{\x \x}]\right\rvert 
        \\ 
        & \qquad
        \leq \frac{C}{\lambda^{n+1}}
        \sum_{N = 0}^{\lvert \Lambda'\rvert} \frac{1}{N!}\left(\frac{C}{2\lambda}\right)^{N}
            \left( \sup_{\substack{Y\subset \Lambda:\\ \lvert Y \rvert = N+1}}
            \int_{R_{+}^{Y}} \d{r}\,  \big|\nu_{Y}(r^{2})\big| \big(\sum_{y \in Y} r_{y}^{2} \big)^{n} \,\prod_{x\in Y} \big(1 +r_x^{2A+1}\big)
            \right) \nonumber
            \\
            &
            \hspace{5cm}
            \times\sum_{\substack{y_1,\ldots,y_N\in\Lambda'\\ \text{ distinct}}}
		\;
		\sum_{T\text{ on }\{1,\ldots, N+1\}} \left(\prod_{(m,n)\in T} P_{y_m y_n} \right). \nonumber
\end{align}
for some constant $C$ depending on $H_{0}$ only.

Using that $V$ satisfies integrable bounds up to order $2(A +n) + 1$, there exists some constant $K_{n}$ depending also on  $A$ such that
\begin{equation}\label{eq:int-bound-proof}
\begin{split}
            \int_{\R_{+}^{Y}} \d{r}\,  & \nu_{Y}(r^{2}) 
           \big(\sum_{y \in Y} r_{y}^{2} \big)^{n} \,\prod_{x\in Y} \big(1 +r_x^{2A+1}\big)
            \leq K_{n}^{|Y|}.            
\end{split}
\end{equation}
Furthermore, standard tree-stripping estimates, see, e.g., \cite{Brydges1984}, give for fixed $T$
\begin{equation}\label{eq:tree-stripping}
\sum_{\substack{y_1,\ldots,y_N\in\Lambda'\\ \text{ distinct}}}
		\prod_{(m,m')\in T} \lvert P_{y_m y_{m'}}\rvert
		\leq \Big( \sup_{x} \sum_{y} |P_{x y}|\Big)^{N} \leq C^{N} \;.
\end{equation}
Thus, since by Cayley's theorem the number of trees on $N+1$ unlabelled vertices is $(N+1)^{N-1}\leq N! \ee^{N+1}$, plugging \eqref{eq:int-bound-proof} and \eqref{eq:tree-stripping} into \eqref{eq: semifinal-G-bound} gives
\[
\left\lvert  \partial^{n}_{E}\E[(G_\Lambda(E+ \ii\varepsilon))_{\x \x}]\right\rvert  \leq  \frac{C'_{n}}{\lambda^{n+1}}\sum_{N=0}^{\lvert \Lambda'\rvert}\left(\frac{C_{n}}{\lambda}\right)^N \;.
\]
The claim follows for $\lambda$ large enough, depending on $H_{0}$ and on the constant $K_{n}$ from \eqref{eq:int-bound-proof}.
\end{proof}

\appendix 

\section{Supersymmetric formalism}
\label{appendix:GA}
 
This appendix summarizes the elements of the supersymmetric formalism used throughout the paper. Standard references are \cite{Berezin1987,Wegner2016}. 

\subsection{Grassmann calculus}
\label{sec: app-grass}

A Grassmann algebra is a complex unital algebra whose generators anticommute. We let 
\[ 
\GA_\Lambda =\GA[ (\xi_x, \bar \xi _{x})_{x \in \Lambda}] 
\] 
denote the Grassmann algebra generated by $(\xi_{x},\bar \xi_{x})_{ x \in \Lambda }$, for some finite set $\Lambda$. Anticommuting generators means that they are subject to the relations
\begin{equation*}
\{\xi_{x} , \xi_{y} \} = \{\xi_{x} , \bar\xi_{y} \} = \{\bar \xi_{x} , \bar\xi_{y} \} = 0 \qquad \quad \forall   x,y \in \Lambda \;,
\end{equation*}
where $\{\cdot , \cdot \}$ denotes the anticommutator. 
Every element $a\in\mathcal{A}_\Lambda$ admits a unique expansion
\[
a = \sum_{I,I'} a_{I,I'} \,\xi_I \bar \xi_{I'} ,
\]
where $I$ and $I'$ range over ordered multi-indices in $\Lambda$, $a_{I,I'} \in\C$, while $\xi_{I}$ and $\bar \xi_{I'}$ denote the associated ordered monomials, with the convention $\xi_{\emptyset} = \bar \xi_{\emptyset} = 1$. 

The parity automorphism $\mathscr P(\xi_i)=-\xi_i$, $\mathscr P(\bar\xi_i)=-\bar\xi_i,$ induces the decomposition
\[
\GA_\Lambda=\GA_\Lambda^0\oplus\GA_\Lambda^1
\]
into even and odd elements. Note that elements in $\mathcal{A}_\Lambda^0$ commute and elements in $\mathcal{A}_\Lambda^1$ anticommute. We write
\[
a=b_a+n_a,
\]
where $b_a:=a_{\emptyset}\in\C$ is the body of $a$, and $n_{a}$ is its nilpotent part.

If $a\in (\GA_\Lambda^0)^k$, then $a=b_a+n_a$, where $b_a\in\R^k$ and $n_a$ is nilpotent. Consequently, for any $f\in C^m(\R^k)$ with $m\ge |\Lambda|$, we may define
\begin{equation}\label{definition:taylor_extension}
f(a) := \sum_{\alpha\in\mathbb N_0^k}
\frac{\partial^\alpha f(b_a)}{\alpha!}
\,n_a^\alpha \;.
\end{equation}
Since $n_a$ is nilpotent, the above Taylor expansion is finite and therefore well-defined.

It is convenient to equip the Grassmann algebra with an $\ell^1$-type norm. For any $a \in \GA_{\Lambda}$, with decomposition $a = \sum_{I,I'} a_{I,I'} \,\xi_I \bar \xi_{I'}$, we define
\begin{equation}\label{eq:def-l1-norm}
\|a \|:=\sum_{I,I'} |a_{I,I'}|\;.
\end{equation}
One can easily check that
\begin{equation*}\label{eq:product-norm}
\lVert ab\rVert \leq \lVert a\rVert \lVert b\rVert \;, \qquad \forall a, b\in \mathcal{A} \;,
\end{equation*}
and $\lVert 1 \rVert= 1$, so that $(\mathcal{A}_{\Lambda}, \lVert \,\cdot\,\rVert)$ is a normed algebra, see \cite[Lemma 2.4]{Fresta2021} for details.
    
We now recall Grassmann derivatives. For given $x\in \Lambda$, any $a\in\mathcal{A}_\Lambda$ has the unique decomposition
	\[a = a_1 + \xi_x a_2 \]
for some $a_1, a_2 \in \GA_\Lambda$ independent of $\xi_x$. We define the (left) derivative by
	\[
	\partial_{\xi_x} a: = a_2 \;.
	\]
A similar definition holds for $\partial_{\bar \xi_x}$. One can see that
\[
\{\partial_{\xi_x}, \partial_{\xi_y}\} = \{\partial_{\xi_x}, \partial_{\bar \xi_y}\} =\{\partial_{\bar \xi_x}, \partial_{\bar \xi_y}\} = 0 \;, \qquad \forall x,y \in \Lambda
\]
Also, one can prove a product rule:
\begin{align}\label{eq:product_rule_for_diff_on_grassmann_algebra}
		&\partial_{\xi_x}(ab)=\left(\partial_{\xi_x}a\right)b + \mathscr{P}(a)\left(\partial_{\xi_x}b\right) \;,
\end{align}
where $\mathscr{P}$ is the parity automorphism introduced above. 
Grasmmann integration coincides with the derivative; the flat Berezin integration form is 
\begin{align*}
		\int \d{\xi_{\Lambda}} a(\xi) = \Big[\prod_{x \in \Lambda}\partial_{\bar \xi_x} \partial_{ \xi_x} \Big] a(\xi) \;,
\end{align*}
where the order of the pairs $\partial_{\bar \xi_x} \partial_{ \xi_x} $ is irrelevant because they commute.
	
The norm introduced above controls Berezin integration through
\begin{equation}\label{ineq:norm}
\left\lvert\int \d{\xi_\Lambda} a(\xi) \right\rvert  \leq \lVert a\rVert \;.
\end{equation}

Since quadratic expressions in Grassmann variables are even and nilpotent, the exponential $\ee^{-\sum_{x,y \in \Lambda}\bar\xi_x M_{xy} \xi_y}$ is a finite polynomial and therefore well-defined.

This allows one to define Gaussian Grassmann integrals in close analogy with ordinary Gaussian integration. The resulting identities are, however, notably different due to the anticommutativity of Grassmann variables.
\begin{proposition}[Gaussian Integrals over Grassmann Variables]\label{prop:grassmann-gaussian} 
Consider the Grassmann algebra $\mathcal{A}_{\Lambda}$. For any matrix $M\in\C^{\Lambda \times \Lambda}$ it holds
\[
\int \d{\xi_\Lambda}\, \ee^{-\sum_{x,y \in \Lambda}\ov{\xi}_x M_{xy}\xi_y} = \det M \;.
\]
Furthermore, if $M$ is invertible, the following integration-by-part formulas holds for any $a \in \GA_{\Lambda}$
\begin{equation}\label{eq:Wick}
\begin{split}
\int \d{\xi_\Lambda}\, \ee^{-\sum_{x,y \in \Lambda}\ov{\xi}_x M_{xy}\xi_y}  \xi_{x} a & =  \int \d{\xi_\Lambda}\, \ee^{-\sum_{x,y \in \Lambda}\ov{\xi}_x M_{xy}\xi_y}  \sum_{y \in \Lambda} M^{-1}_{xy} \partial_{\bar \xi_{y}} a
\\
\int \d{\xi_\Lambda}\, \ee^{-\sum_{x,y \in \Lambda}\ov{\xi}_x M_{xy}\xi_y}  \ov{\xi}_{x} a & =  -\int \d{\xi_\Lambda}\, \ee^{-\sum_{x,y \in \Lambda}\ov{\xi}_x M_{xy} \xi_y}  \sum_{y \in \Lambda} M^{-1}_{xy} \partial_{ \xi_{y}} a
\end{split}
\end{equation}
\end{proposition}
The proof is combinatorial and can be found in
\cite[Section 3.3]{Wegner2016}. The determinant appearing in
Proposition~\ref{prop:grassmann-gaussian} is precisely the feature that makes Grassmann variables useful in the study of random operators. Combined with ordinary Gaussian integrals, it allows for exact cancellations between determinants and inverse determinants, which is at the heart of the supersymmetric trick.
    	
\subsection{Supersymmetric trick} 

We introduce supervectors of the form
\begin{equation*}
\Phi_{x} = (\phi_{x},\xi_{x})^{t} \;, \qquad  \Phi^{*}_{x}= (\ov{\varphi}_{x},\ov{\xi}_{x}) \;,
\end{equation*}
where $\cdot^{t}$ denotes transposition, consisting of complex variables $\phi_{x}  \in \C$, $\bar \phi_{x}$ denoting the complex conjugate, and Grassmann variables $\xi_{x}, \bar \xi_{x} \in \GA_{\Lambda}$. We denote $\Phi = (\Phi_{x})_{x \in \Lambda}$ and $ \Phi^* = ( \Phi^*_{x})_{x \in \Lambda}$, and use the conventions
\[
\Phi^*_x\Phi_y = \bar\phi_x \phi_{y}+ \ov{\xi}_x \xi_y \;, \qquad \forall x,y \in \Lambda
\]
together with
\[
\Phi^*\Phi = \sum_{x\in \Lambda} \Phi^*_x\Phi_x  \quad  \qquad
\Phi^* M\Phi :=  \sum_{x,y \in \Lambda} M_{xy}\Phi^*_x\Phi_y , \qquad 	M\in\C^{\Lambda\times \Lambda} \;.
\]
We also introduce the superintegration forms
\[
\dPhi{x} = \frac{\d \mathrm{Re} \phi_{x}\d \mathrm{Im} \phi_{x} }{\pi}\partial_{\ov{\xi}_x}\partial_{\xi_x}\;,\qquad \dPhi{\Lambda} := \prod_{x \in \Lambda} \dPhi{x}\;.
\]
The cancellation between the determinants from standard and Grassmann integrations leads to the following representation formula, which is the starting point of the supersymmetric method.
\begin{lemma}\label{lemma:SUSY_Gauß_int}
For any complex symmetric matrix $M \in \C^{\Lambda \times \Lambda}$ with positive definite real part we have
\begin{equation}\label{eq:susy-gaussian-integr}
1 = \int \dPhi{\Lambda}  \,\ee^{-\Phi^* M\Phi}
\;, \qquad 
M^{-1}_{xy} = \int \dPhi{\Lambda} \xi_x \ov{\xi}_y \,\ee^{-\Phi^* M\Phi} \;.
\end{equation}
\end{lemma}
This is a direct consequence of Proposition~\ref{prop:grassmann-gaussian} together with standard Gaussian integration.

The first identity in \eqref{eq:susy-gaussian-integr} is an example of a supersymmetric cancellation.
To describe these cancellations more generally, we introduce the supersymmetry generator acting on functions of the supervector $\Phi$.
The supersymmetry generator is the following odd differential operator
\[
		Q_{\Phi} = \sum_{x \in \Lambda}
\Big[ 
		 \bar\xi_{x}\partial_{\bar \phi_x} - \,\bar\phi_x \partial_{ \bar \xi_x}
		 + \xi_x\partial_{\phi_x} + \,\phi_x \partial_{\xi_x}	
		 \Big]			 \;.
		 \]
A function $f$ is supersymmetric if it is $Q$-closed, i.e. if it is differentiable and 
		\[Q_\Phi f(\Phi) = 0.\]

For example, the following class of functions is supersymmetric.
\begin{lemma}\label{lemma:conditions_on_f_to_be_supersymmetric}
Let $\mathcal{P}_{\Lambda}$ be the set of unordered pairs in $\Lambda$ and let $f\colon\, \R^{\mathcal{P}_\Lambda}\rightarrow\C$ be a smooth function. Then
\[
f(\{\Phi^*_x\Phi_y+\Phi^*_y\Phi_x\}_{(xy) \in \mathcal{P}_{\Lambda}})
 \]
is a well-defined supersymmetric function.
\end{lemma}
\begin{proof}
 With slight abuse of notation, we abridge $f(\Phi) \equiv f(\{\Phi^*_x\Phi_y+\Phi^*_y\Phi_x\}_{(xy) \in \mathcal{P}_{\Lambda}})$.
First of all, note that the body of $\Phi^*_x\Phi_y+\Phi^*_y\Phi_x$ is real, hence $f(\Phi)$ makes sense by \eqref{definition:taylor_extension}.
The chain rule holds by \eqref{definition:taylor_extension} and with \eqref{eq:product_rule_for_diff_on_grassmann_algebra} implies
\begin{align*}
\frac{\partial}{\partial \xi_x}f(\Phi)
        &= \sum_{(yy') \in \mathcal{P}_{\Lambda}} \partial_{(y y')}f(\Phi)  \partial_{\xi_x}(\Phi^*_y\Phi_{y'}+\Phi^*_{y'}\Phi_y)
        =-\sum_{y} \ov{\xi}_y \partial_{(y x)} f(\Phi) \;,
\end{align*}
where $\partial_{(y y')}$ denote the derivative with respect to the coordinate associated with the unordered pair $(y y')$.
Similarly, we can compute the derivatives with respect to $\ov{\xi}_j, \phi_j$ and $\ov{\phi}_j$.  Using these computations we see that 
\begin{equation*}
\begin{split}
Q_{\Phi}f(\Phi) & =\sum_{(xy) \in \mathcal{P}_\Lambda}
\big(Q_{\Phi} (\Phi^*_x\Phi_y) + Q_{\Phi} ( \Phi^*_y\Phi_x )\big)	\partial_{(yx)}f(\Phi)  =0
\end{split}
\end{equation*}
where we used that $Q_{\Phi} ( \Phi^*_x\Phi_y )= \ov{\xi}_x\phi_y-\ov{\phi}_x\xi_y+\xi_x\ov{\phi}_y-\phi_x\ov{\xi}_y$.
\end{proof}

Most cancellation mechanisms used in the supersymmetric method ultimately stem from the following remarkable identity.
\begin{theorem}[SUSY localisation formula]	\label{theorem:SUSY_Localization}
Let $f$ be supersymmetric Schwartz function, then
		\[\int \dPhi{\Lambda} f(\Phi) = f(0).\]
\end{theorem}
A proof can be found in \cite[Appendix C]{Disertori2010}. A general exposition of localisation results for integrals over supermanifolds can be found in \cite{Schwarz1997}.   A simple consequence of the SUSY localisation formula is the identity
\begin{equation*}
\int \dPhi{\Lambda} \xi_x \ov{\xi}_y \,\ee^{-\Phi^* M\Phi}
= \int \dPhi{\Lambda} \phi_x \ov{\phi}_y \,\ee^{-\Phi^* M\Phi} \;.
\end{equation*}
which illustrates the correspondence between complex and Grassmann correlation functions generated by supersymmetry.

We conclude this section with an estimate that is repeatedly used throughout the paper. 
\begin{lemma}\label{lemma:bound-norm}
For any superfunction $f(\Phi)$, we have the bound
\begin{align*}
    	\left\lvert \int \dPhi{\Lambda} \,f(\Phi) \right\rvert \leq \int \dphi{\Lambda} \lVert f(\Phi)\rVert \;.
\end{align*}
\end{lemma}
This is a direct consequence of \eqref{ineq:norm}.

\paragraph{Acknowledgements} 
The research at the University of Bonn was supported by the German Research Foundation (DFG) under Germany’s Excellence Strategy-GZ 2047/1 - 390685813, and by the Deutsche Forschungsgemeinschaft (DFG, German Research Foundation) - CRC 1720 - 539309657.
L.~F.~acknowledges partial support by the European Union’s Horizon 2020 research and innovation programme under the Marie Sk\l odowska-Curie grant agreement No 101154394.

\paragraph{AI Statement} The mathematical content of this work was developed without AI assistance. Free AI-based tools were partly used for assistance with English grammar. The authors take full responsibility for the content of the manuscript.

\bibliographystyle{plainnat}

\begin{thebibliography}{10}

\bibitem{Anderson79}
E.~Abrahams, P.~W. Anderson, D.C. Licciardello, and T.V. Ramakrishnan.
\newblock Scaling theory of localization: absence of quantum diffusion in two
  dimensions.
\newblock {\em Phys. Rev. Lett.}, 42:673, March 1979.

\bibitem{Aiz93}
M.~Aizenman and S.~Molchanov.
\newblock Localization at large disorder and at extreme energies: an elementary
  derivation.
\newblock {\em Commun. Math. Phys.}, 157:245--278, 1993.

\bibitem{aizenman2015random}
M.~Aizenman and S.~Warzel.
\newblock {\em Random Operators: Disorder Effects on Quantum Spectra and
  Dynamics}.
\newblock Graduate Studies in Mathematics. American Mathematical Society, 2015.

\bibitem{Anderson1958}
P.~W. Anderson.
\newblock Absence of diffusion in certain random lattices.
\newblock {\em Physical Review}, 109(5):1492–1505, March 1958.

\bibitem{AFP20}
G.~Antinucci, L.~Fresta, and M.~Porta.
\newblock A supersymmetric hierarchical model for weakly disordered 3d semimetals.
\newblock {\em Ann. Henri Poincar\'e}, 21(11):3499--3574, 2020.

\bibitem{Bao18}
Z.~Bao and E.~Erd\H{o}s.
\newblock Delocalization for a class of random block band matrices.
\newblock {\em Probab. Theory Relat. Fields}, 167:673--776, 2018.

\bibitem{Berezin1987}
F.A. Berezin.
\newblock {\em Introduction to Superanalysis}.
\newblock Mathematical Physics and Applied Mathematics. Springer, 1987.

\bibitem{Bourg}
J.~Bourgain and C.E. Kenig.
\newblock On localization in the continuous Anderson-Bernoulli model in higher dimension.
\newblock {\em Invent. Math.}, 161(2):389--426, 2005.

\bibitem{Bovier1990}
A.~Bovier.
\newblock The density of states in the Anderson model at weak disorder: A renormalization group analysis of the hierarchical model.
\newblock {\em Journal of Statistical Physics}, 59(3–4):745–779, May 1990.

\bibitem{Bovier1988}
A.~Bovier, A.~Campanino, A.~Klein, and J.~F. Perez.
\newblock Smoothness of the density of states in the Anderson model at high   disorder.
\newblock {\em Communications in Mathematical Physics}, 114(3):439–461,
  September 1988.

\bibitem{Brydges1984}
David~C. Brydges.
\newblock A shourt course on cluster expansions.
\newblock Les Houches, Session XLIII, Critical phenomena, random symstems,
  gauge theories, 1984.

\bibitem{klein86-}
M.~Campanino and A.~Klein.
\newblock A supersymmetric transfer matrix and differentiability of the density   of states in the one-dimensional Anderson model.
\newblock {\em Commun. Math. Phys.}, 104:227--241, 1986.

\bibitem{Carmona1987}
R.~Carmona, A.~Klein, and F.~Martinelli.
\newblock Anderson localization for Bernoulli and other singular potentials.
\newblock {\em Commun. Math. Phys.}, 108(3):41--66, 1987.

\bibitem{Carmona1990}
R.~Carmona and J.~Lacroix.
\newblock {\em Spectral Theory of Random Schr\"{o}dinger Operators}.
\newblock Birkh\"{a}user Boston, 1990.

\bibitem{Combes98}
J.~M. Combes, P.~D. Hislop, and E.~Mourre.
\newblock Correlated Wegner inequalities for random Schrödinger operators.
\newblock {\em n: Advances in Differential Equations and Mathematical Physics},
  pages 191--203, 1998.

\bibitem{Constantinescu1984}
F.~Constantinescu, J.~Fr{\"o}hlich, and T.~Spencer.
\newblock Analyticity of the density of states and replica method for random   Schr\"odinger operators on a lattice.
\newblock {\em J. Stat. Phys.}, 34(3–4):571–596, February 1984.

\bibitem{Craig1983}
W.~Craig and B.~Simon.
\newblock Log {H\"o}lder continuity of the integrated density of states for stochastic {J}acobi matrices.
\newblock {\em Commun.Math. Phys.}, 90(2):207--218, 1983.

\bibitem{Delyon1984}
F.~Delyon and B.~Souillard.
\newblock Remark on the continuity of the density of states of ergodic finite difference operators.
\newblock {\em Commun.Math. Phys.}, 94(2):289–291, June 1984.

\bibitem{Ding20}
J.~Ding and C.~K. Smart.
\newblock Localization near the edge for the Anderson Bernoulli model on the two dimensional lattice.
\newblock {\em Invent. math.}, 219:467--506, 2020.

\bibitem{Disertori2020}
M.~Disertori and M.~Lager.
\newblock Supersymmetric polar coordinates with applications to the Lloyd model.
\newblock {\em Mathematical Physics, Analysis and Geometry}, 23(1), January
  2020.

\bibitem{Disertori2010+}
M.~Disertori and T.~Spencer.
\newblock Anderson localization for a supersymmetric sigma model.
\newblock {\em Commun.Math. Phys.}, 300(3):659--671, 2010.

\bibitem{Disertori2010}
M.~Disertori, T.~Spencer, and M.~R. Zirnbauer.
\newblock Quasi-diffusion in a 3d supersymmetric hyperbolic sigma model.
\newblock {\em Commun.Math. Phys.}, 300(2):435–486, August 2010.

\bibitem{Efetov1980}
K.~Efetov, A.I. Larkin, and D.E. Khmelnitskii.
\newblock Interaction of diffusion modes in the theory of localization.
\newblock {\em Journal of experimental and theoretical physics}, 52:568–574,
  1980.

\bibitem{Elgart}
A.~Elgart.
\newblock Lifshitz tails and localization in the three-dimensional Anderson model.
\newblock {\em Duke Math. J.}, 146:331--360, 2009.

\bibitem{Erdos25}
L.~Erdős and V.~Riabov.
\newblock The zigzag strategy for random band matrices.
\newblock {\em Preprint}, arXiv:2506.06441, 2025.

\bibitem{Fresta2021}
L.~Fresta.
\newblock Supersymmetric cluster expansions and applications to random {S}chr\"{o}dinger operators.
\newblock {\em Mathematical Physics, Analysis and Geometry}, 24(1), 2021.

\bibitem{Froe07}
R.~Froese, D.~Hasler, and W.~Spitzer.
\newblock Absolutely continuous spectrum for the Anderson model on a tree: a geometric proof of Klein’s theorem.
\newblock {\em Commun. Math. Phys.}, 269(1):239--257, 2007.

\bibitem{Fro85}
J.~Fr\"ohlich, F.~Martinelli, E.~Scoppola, and T.~Spencer.
\newblock Constructive proof of localization in the Anderson tight binding model.
\newblock {\em Commun. Math. Phys.}, 101:21--46, 1985.

\bibitem{Fro83}
J.~Fr\"ohlich and T.~Spencer.
\newblock Absence of diffusion in the Anderson tight binding model for large
  disorder or low energy.
\newblock {\em Commun. Math. Phys.}, 88:151--184, 1983.

\bibitem{KirschInvitation2007}
W.~Kirsch.
\newblock An invitation to random {S}chr{\"o}dinger operators, 2007.

\bibitem{Klein}
A.~Klein.
\newblock Absolutely continuous spectrum in the Anderson model on the Bethe lattice.
\newblock {\em Math.~Res.~Lett.}, 1:399--407, 1994.

\bibitem{klein86}
A.~Klein, F.~Martinelli, and J.F. Perez.
\newblock A rigorous replica trick approach to Anderson localization in one dimension.
\newblock {\em Commun. Math. Phys.}, 106:623--633, 1986.

\bibitem{Klopp}
F.~Klopp.
\newblock Weak disorder localization and Lifshitz tails.
\newblock {\em Commun. Math. Phys.}, 232:125--155, 2002.

\bibitem{Figotin1992}
L.~Pastur and A.~Figotin.
\newblock {\em Spectra of Random and Almost-Periodic Operators}.
\newblock Springer Berlin, Heidelberg, 1992.

\bibitem{Rudin}
W.~Rudin.
\newblock {\em Real and Complex Analysis}.
\newblock McGraw-Hill, 1987.

\bibitem{Schwarz1997}
A.~Schwarz and O.~Zaboronsky.
\newblock Supersymmetry and localization.
\newblock {\em Communications in Mathematical Physics}, 183(2):463--476, January 1997.

\bibitem{Shcherbina21}
M.~Shcherbina and T.~Shcherbina.
\newblock Universality for 1d random band matrices.
\newblock {\em Commun. Math. Phys.}, 385:667--716, 2021.

\bibitem{Sjostrand1999-1}
J~Sjostrand and W~Wang.
\newblock Supersymmetric measures and maximum principles in the complex domain.
  exponential decay of Green’s functions.
\newblock {\em Annales Scientifiques de l’École Normale Supérieure},
  32(3):347–414, May 1999.

\bibitem{Wang}
W.-M. Wang.
\newblock Localization and universality of Poisson statistics for the  multidimensional Anderson model at weak disorder.
\newblock {\em Invent. math.}, 146:365--398, 2001.

\bibitem{Wegner2016}
F.~Wegner.
\newblock {\em Supermathematics and its Applications in Statistical Physics}.
\newblock Springer Berlin Heidelberg, 2016.

\bibitem{Wegner1979}
F.J. Wegner.
\newblock The mobility edge problem: Continuous symmetry and a conjecture.
\newblock {\em Zeitschrift f{\"u}r Physik B Condensed Matter and Quanta},
  35(3):207--210, September 1979.

\bibitem{Yau25}
H.-T. Yau and J.~Yin.
\newblock Delocalization of one-dimensional random band matrices.
\newblock {\em Preprint}, 2025.

\end{thebibliography}

\begingroup
\footnotesize

\end{document}